\documentclass[lettersize,journal]{IEEEtran}
\usepackage{amsmath,amsfonts}
\usepackage{algorithmic}
\usepackage{algorithm}
\usepackage{array}
\usepackage{balance}
\usepackage{flushend}
\usepackage[caption=false,font=normalsize,labelfont=sf,textfont=sf]{subfig}
\usepackage{textcomp}
\usepackage{stfloats}
\usepackage{url}
\usepackage{verbatim}
\usepackage{graphicx}
\usepackage{cite}
\usepackage{listings}
\usepackage{amsmath,amsfonts}
\usepackage{graphicx}
\usepackage{amsfonts}
\usepackage{array}
\usepackage{amssymb}
\usepackage{helvet}
\usepackage{color}
\usepackage{bm}
\usepackage{float}
\usepackage{cite}
\usepackage{multirow}
\usepackage{amssymb}
\usepackage[caption=false,font=normalsize,labelfont=sf,textfont=sf]{subfig}
\usepackage{amsmath,amsfonts}
\usepackage{algorithmic}
\usepackage{algorithm}
\usepackage{textcomp}
\usepackage{stfloats}
\usepackage{url}
\usepackage{verbatim}
\usepackage{graphicx}
\usepackage{cite}
\usepackage{amssymb}
\usepackage{makecell}
\usepackage{tabularx}
\usepackage{booktabs} % 用于三线表格
\usepackage{array}    % 用于更灵活的列格式
\newcommand{\RNum}[1]{\uppercase\expandafter{\romannumeral #1\relax}}

\begin{document}

	\title{ Bridge Micro-Deformation Monitoring Scheme with Integrated Sensing and Communications}
	\author{Boxuan Sun, Hongliang Luo, Shaodan Ma, and Feifei Gao
		
		\thanks{Boxuan Sun, Hongliang Luo, and Feifei Gao are with 
			Department of Automation, Tsinghua University, 
			Beijing 100084, China (email: sunbx26@163.com; luohl23@mails.tsinghua.edu.cn; feifeigao@ieee.org).
			
			Shaodan Ma is with the State Key Laboratory of Internet of Things for Smart City and the Department of Electrical and Computer Engineering, University of Macau, Macao S.A.R. 999078, China (e-mail: shaodanma@um.edu.mo).
			
			}
		% <-this % stops a space
		%\thanks{Manuscript received April 19, 2021; revised August 16, 2021.}
	}
	
	
	
	\maketitle

	\begin{abstract}
		In this paper, we propose a novel integrated sensing and communications (ISAC) scheme to perform bridge micro-deformation monitoring (BMDM) in complex environments. 
		We first provide an excitation-bridge coupling model to represent the micro-deformation process of the bridge.
		Next, we design a novel frame structure for BMDM applications, and construct the OFDM echo channel model for basic scene of BMDM, including micro-deformation, dynamic objects, and static environment. 
		Then,  we develop a phasor statistical analysis method based on average cancellation algorithm to suppress the interference of dynamic objects, as well as a circle fitting method based on least squares algorithm to remove the interference of  static environment near the monitoring area.
		Furthermore, we extract the micro-deformation feature vector from the OFDM echo signals after inverse discrete fourier transform (IDFT), and derive vertical micro-deformation value with the time-frequency phase resources.
		Simulation results demonstrate the effectiveness of the proposed BMDM scheme and its robustness against both dynamic interferences and static interferences.
		
	\end{abstract}


	\section{Introduction}
	In recent years, structural health monitoring (SHM) has gained widespread applications in smart city development\cite{zonzini2020structural,di2021structural,dutta2021recent}. The idea is to ensure the safety and integrity of critical infrastructure such as mines, tunnels, high-rise buildings, and bridges. SHM employs advanced sensor technology to progressively conduct crack exploration, micro-deformation monitoring, and stress-strain analysis in real time on these structures\cite{zonzini2020vibration,sarwar2020multimetric,9547763}. Among the broad of monitoring tasks, bridge micro-deformation monitoring (BMDM) is of particularly importance.
	
	BMDM captures and analyzes subtle deformations in critical bridge components that includes decks, columns, piers, cables, steel beam connections, and foundation settlement points\cite{liu2020internet,yang2022research,shi2024board}. These components are susceptible to minor changes caused by factors such as load variations, temperature fluctuations, and long-term fatigue.  In case these  micro-deformations are overlooked, severe consequences like traffic disruptions, structural failures, and even catastrophic bridge collapses may arise, which threaten public safety and social stability\cite{zinno2022artificial,fawad2024integration}.

	One type of BMDM utilizes contact sensors such as vibrometers, accelerometers, strain gauges and displacement meters on bridges\cite{zalt2007evaluating,bruschetta2013fusion}. These sensors evaluate the dynamic response and health status of bridges by combining vibration frequency, stress change and displacement data with time domain or frequency domain analysis methods such as fast fourier transform (FFT)\cite{koganezawa2024vibration,zhou2024novel,ferguson2024systematic}. However, contact sensors have to be installed in fixed positions and are difficult to move or adjust, which limits the monitoring range and can only cover a few pre-selected points on the bridges.

	Another type of BMDM employs non-contact sensors\cite{michel2023assessing,pramudita2023fmcw,sun2024deformation}. For example, interferometric synthetic aperture radar (InSAR) and global navigation satellite system (GNSS) can remotely detect bridge micro-deformation by emitting electromagnetic waves and analyzing reflected signals\cite{wang20242,wang2021review,civera2021computer}. These technologies typically achieve centimeter to millimeter-level precision. Compared with contact monitoring methods, non-contact monitoring can flexibly monitor multiple parts, and can even cover the entire bridge structure. These non-contact characteristics avoid disrupting traffic and are suitable for large bridges and complex environments\cite{alonso2024contribution,hu2024vehicle,topal2024filtering}. However, these monitoring systems require separate deployment and regular maintenance, such as setting up special radar equipment, adjusting observation angles, and ensuring signal stability, which increases the cost of operation and maintenance.
	
	The recently developed integrated sensing and communications (ISAC) in the coming 6G technology can provide an alternative solution for BMDM\cite{liu2022integrated,giordani2020toward,de2021convergent}. In ISAC system, the base station (BS) can utilize the orthogonal frequency division multiplexing (OFDM) signal to achieve the sensing task of target area\cite{zhong2025resource,2024arXiv240519925L,liao2023optimized}, without the need of additional independent monitoring devices. For example, \cite{yang2024application} introduces the application of reconfigurable intelligent surfaces (RIS) as reference points, which can dynamically adjust phases to generate varying radio signals and realize millimeter-level micro-deformation monitoring through Fisher information analysis. In \cite{10868647}, the authors propose an ISAC-based deformation monitoring method that employs widely deployed 5G BSs to transmit configured signals and achieves high-precision detection by analyzing real-time phase variations in echo signals from a displaced corner reflector that simulates micro-deformations in ranging-angle maps.
	However, these studies on phase-based micro-deformation monitoring overlook the limited range of phase (i.e., the problem of phase wrapping), the impact of the broadband characteristics of OFDM systems on phase, and the handling of dynamic objects and static clutter near the monitoring area.

	In this paper, we propose a novel ISAC scheme to perform BMDM in complex environments that include dynamic objects and static clutter.
	The main contributions of this paper are summarized as follows:

	\begin{itemize}
		
		\item 
		We provide an excitation-bridge coupling model to represent the micro-deformation process of the birdge and design an novel frame structure for BMDM applications. 
		\item
		We construct the OFDM echo channel model for basic scene of BMDM that includes micro-deformation, dynamic objects, and static environment.
		\item
		We develop a phasor statistical analysis method based on average cancellation algorithm to suppress the interference of dynamic objects on the bridge, and a circle fitting method based on least squares algorithm to remove the interference of  static environment.
		
		\item 
		We extract the micro-deformation feature vector from the OFDM echo signals, and derive the vertical micro-deformation value of the bridge deck by calculating the multi-frame phase difference.
		To address the limited range of phase, we present a phase unwrapping method based on historical micro-deformation data.
		\item 
		Simulation results demonstrate the effectiveness of the proposed scheme and its robustness against both dynamic interferences and static interferences.
	\end{itemize}
	
	The remainder of this paper is organized as follows.
	In Section \RNum{2}, we design an OFDM frame structure and construct an echo channel model for basic scene of BMDM .
	In Section \RNum{3}, we utilize clutter suppression methods to suppress dynamic interferences and static interferences.
	In Section \RNum{4}, we derive the vertical micro-deformation value.
	Simulation results and conclusions are given in Section \RNum{5} and Section \RNum{6}.

	\begin{figure}[!t]
		\centering
		\includegraphics[width=80mm]{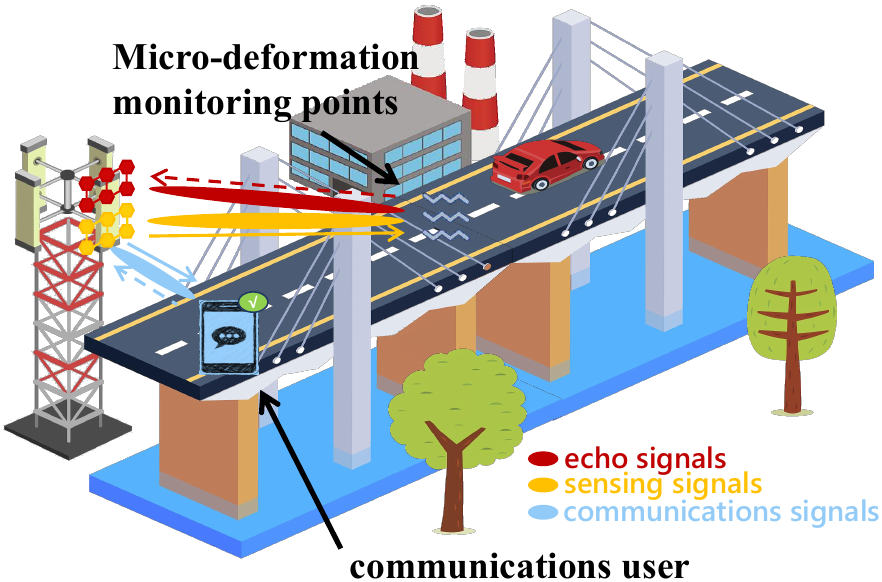}
		\caption{System model.}
		\label{fig_2}
	\end{figure}
	
	\emph{Notation}:
	Lower-case and upper-case boldface letters $\mathbf{a}$ and $\mathbf{A}$ denote a vector and a matrix;
	$\mathbf{a}^{*}$, $\mathbf{a}^{T}$ and $\mathbf{a}^{H}$ denote the conjugate, transpose and conjugate transpose of $\mathbf{a}$, respectively
	$\mathbf{a}[n]$  denotes the $n$-th element of the vector $\mathbf{a}$;
	$\mathbf{A}[i,j]$ denotes the $(i,j)$-th element of the matrix $\mathbf{A}$; $\mathbf{A}[i_1:i_2,:]$ is the submatrix composed of all columns elements in rows $i_1$ to $i_2$ of matrix $\mathbf{A}$;
	$\mathbf{A}[:,j_1:j_2]$ is the submatrix composed of all rows elements in columns $j_1$ to $j_2$ of matrix $\mathbf{A}$;
	$\left|\cdot\right|$  denotes the absolute operator;
	$\| \cdot \|$ denotes the euclidean distance;
	$\left \lfloor \cdot \right \rceil$ represents rounding to the nearest integer;
	$\otimes$ denotes the Kronecker product;
	$\odot$ represents element wise multiplication;
	$\mathbb{Z}$ denotes the integer set;
	$\mathbb{R}$ and $\mathbb{C}$ represent the real field and complex field, respectively;
	$\mathbb{R}^{q}$ represents $q$-dimensional real vectors;
	$\mathbb{R}^{+}$ represents positive real numbers;
	$ \mathbb{E}{·}$ is the mathematical expectation operator. In addition, some of the key notations are listed in Table I.
	
	%$\mathcal{CN}(\mu,\sigma^2)$ denotes the Gaussian distribution with mean $\mu$ and covariance $\sigma^2$.
	%And $\mathcal{U}(a,b)$ denotes the uniform distribution between $a$ and  $b$.%
	\begin{table*}[h!]
		\centering
		\color{black}
		\caption{List of Key Notations}
		\begin{tabular}{cc}
			\toprule[1.5pt]
			Notation & Description\\
			\midrule[1pt]
		$N_{H}$  & Number of antennas  in HU-UPA\\
		$N_{R}$  & Number of antennas in RU-UPA\\
		$M$ & Number of subcarriers in OFDM signals\\
		$N$ & Number of OFDM symbols in one frame\\
		$P$ & Total number of radio frames\\
		$E$ & Number of excitation sources\\
		$K$ & Number of dynamic interferences\\
		$S$ & Number of static interferences\\
		$\zeta$ & Damping coefficient \\
		  $\xi$ & Young's modulus of structural material \\
		$I_B$ & Moment of inertia of the cross-section \\
		$\rho_B$ & Mass per unit length of the bridge \\
		 $A_e$ & Amplitude of the $e$-th excitation source \\
		$f_e$ & Frequency of the $e$-th excitation source \\
		
		$R_{p}^{\delta}$ & Radial distance of BMDM point to BS in the $p$-th frame\\        
		$\theta^{\delta}$ & Initial azimuthal angle of BMDM point relative to BS\\
		$\phi^{\delta}$ & Initial elevation angle of BMDM point relative to BS\\
		$\Psi^{\delta}$ & Horizontal spatial-domain direction of BMDM point\\
		$\Omega^{\delta}$ & Pitch spatial-domain direction of BMDM point\\
		
		$R_{p}^{k}$ & Radial distance of the $k$-th dynamic interference to BS in the       
		$p$-th frame\\
		$v_{p}^{k}$ & Velocity of the $k$-th dynamic interference in the $p$-th frame\\        
		
		$R^{s}$ & Radial distance of the $s$-th static interference to BS\\ 
		$\kappa^{\delta}$ & Range bin index containing the micro-deformation point\\        
		      
		$\varphi_{p}$ & Phase of micro-deformation signal in the $p$-th frame\\
		$\Delta \varphi$ & Phase difference for micro-deformation estimation\\
		
		$f_{p}^{k}$ & Doppler frequency of the $k$-th dynamic interference in the $p$-th    
		frame\\
		$\Delta D_{0\to p}$ & Micro-deformation from initial frame to $p$-th frame\\

		$\mathbf{H}_{p,n,m}^{\text{micro}}$ & Micro-deformation echo channel matrix on     the $m$-th subcarrier at the $n$-th symbol in the $p$-th frame\\
		$\mathbf{H}_{p,n,m}^{k}$ & Echo channel matrix of the $k$-th dynamic
		interference on
		the $m$-th subcarrier at the $n$-th symbol in the $p$-th frame\\
		$\mathbf{H}_{n,m}^{s}$ & Echo channel matrix of the $s$-th static interference      
		on the $m$-th subcarrier at the $n$-th symbol\\
		$\mathbf{Y}[p,n,m]$ & Echo signal tensor element at $p$-th frame, $n$-th symbol,    
		$m$-th subcarrier\\
		$\mathbf{F}[p,n,\kappa]$ & IDFT result of echo signal at $p$-th frame, $n$-th       
		symbol, $\kappa$-th range bin\\
		
			\bottomrule[1.5pt]
		\end{tabular}
	\end{table*}

	\section{ System Model}
	
	In this section, we present the basic scenario of BMDM with  ISAC system, and derive the echo channel models for micro-deformations, dynamic interferences, and static interferences.
	\subsection{ISAC Scene and BS Model}
	The ISAC BS employs co-located massive MIMO arrays operating in mmWave frequency band, which is equipped with one hybrid unit uniform planar array (HU-UPA) of size $N_{H} = N_{H}^{x} \times N_{H}^{z}$ and one radar unit uniform planar array (RU-UPA) of size $N_{R} = N_{R}^{x} \times N_{R}^{z}$, as shown in Fig.~1.  Both arrays share identical spatial coordinates, where the HU-UPA simultaneously manages wireless communications and sensing tasks, while the RU-UPA  is responsible for receiving sensing echo signals\cite{8288677,9947033}. The antenna spacing between the antennas along x-axis and z-axis is $d_{x}=d=\frac{\lambda}{2}$ and $d_{z}=d=\frac{\lambda}{2}$, where $\lambda$ is the wavelength.  Each antenna element connects to an independent phase shifter in the hybrid architecture to enable adaptive beam steering, which facilitates concurrent information transfer and BMDM. The precise beam control is driven by $N_{H}^{RF} \ll N_H$ and $N_{R}^{RF}\ll N_R$ radio frequency (RF) chains, respectively. 
	
	Suppose that the ISAC system emits OFDM signals for joint communications and sensing\cite{10147248,8386661,9755276}. 
	The signal utilizes $N$ OFDM symbols, each containing $M$ subcarriers with a frequency spacing of $\Delta f$. Then, the transmission bandwidth is $B = M \Delta f$, and the $m$-th subcarrier frequency is  $f_m = f_c + m \Delta f, \, m=0,1, \ldots, M-1$, where $f_{c}$ is the carrier frequency. Each OFDM symbol has a duration of $T_{\text {sym}}=T_{\text {sym}}^{\prime}+T_{g}$, where $T_{\text {sym}}^{\prime}=\frac{1}{\Delta f}$ is the effective symbol period and $T_{g}$ is the guard interval. 
	
	The signal frame structure is based on 5G New Radio (NR) protocol, with each frame lasting for $T_{f} = $ 10~ms.   In  time division duplex (TDD) mode, there are four types of slots: monitoring (MT) slots to execute micro-deformation sensing, downlink (DL) slots to transmit communications data from BS to users, uplink (UL) slots to receive feedback and information form users to BS, and flexible (S) slots to facilitate mode-switching transitions, with each slot containing $N^{+}$ OFDM symbols as shown in Fig.~2. Assume that the ISAC system has a total of $P$ frames, where each frame includes 1 ``MMMSU'' pulse repetition interval (PRI) to perform BMDM and 3 ``DDDSU'' PRIs to execute communications, with each PRI lasting $T_{c}=$2.5~ms\cite{10625724}. Thus, the total duration of each frame is $T_{f} = $ 10~ms.

	\begin{figure*}[!t]
		\centering
		\includegraphics[width=180mm]{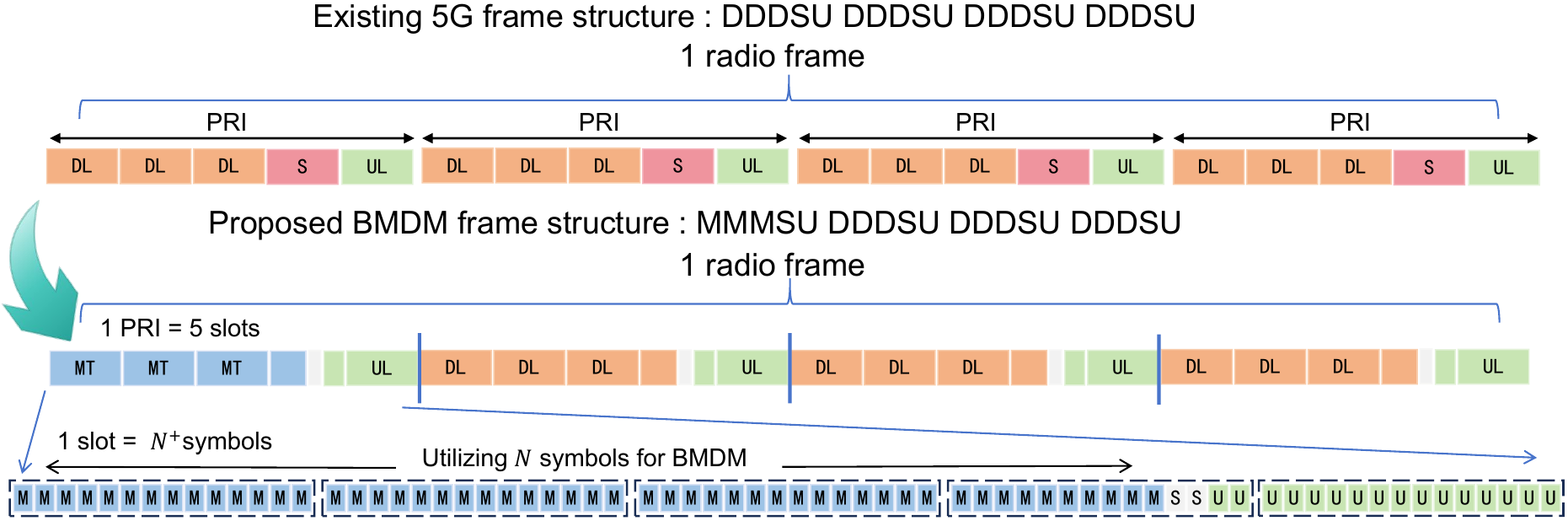}
		\caption{The diagram of the proposed BMDM frame structure.}
		\label{fig_4}
	\end{figure*}

	Let us employ spatial Cartesian coordinates to define positions of BMDM points, dynamic interference sources, and static interference locations, and utilize spherical coordinates  \((\theta, \phi, r)\) for channel analysis. The BS is located at the origin, where \(r > 0\) indicates the radial distance,  the azimuth angle $\theta \in\left[0^{\circ}, 180^{\circ}\right]$, and the elevation angle $\phi \in\left[-90^{\circ}, 90^{\circ}\right]$. The service area of BS is  \(\{(\theta, \phi, r) \mid \theta_{\text{min}} \leq \theta \leq \theta_{\text{max}}, \phi_{\text{min}} \leq \phi \leq \phi_{\text{max}}, r_{\text{min}}\! \leq r \! \leq r_{\text{max}}\}\). We assume that the service area covers all communications users and micro-deformation targets. As the co-design of communications and sensing is well-established in prior studies, we here ignore the communications-stage analysis and focus solely on BMDM during MT slots.
	\subsection{Echo Channel Model of Micro-Deformation}
	
	The establishment of the bridge micro-deformation model should consider various influencing factors.
	
	The free deformation of a bridge is governed by its inherent properties, including material composition and geometric characteristics. This can be understood as the bridge's natural    
	``breathing'' motion that occurs even without external forces. Denote $\xi$ as the elastic modulus of the structural material, $I_{B}$ as the moment of inertia of the cross-section at mid-span, $\rho _{B}$ as the mass per unit of length, and $W$ as the effective span length of the structure, i.e., the length of the bridge between two piers. Define the fundamental frequency of the bridge as $f_{B}=\frac{\pi}{2 W^{2}} \sqrt{\frac{\xi I_{B}}{\rho _{B}}}$. Then, the impact of the free deformation on the bridge can be expressed as\cite{7729680,ma2023structural} 
	\begin{equation}
		\begin{split}\color{black}
			\begin{aligned}
				D_{\text{free}}=A_{0} \sin \left(2 \pi f_{B} t+\varphi_{B}\right),
			\end{aligned}
		\end{split}
	\end{equation}
	where $A_{0}$ and $\varphi _{B}$ are the amplitude and initial phase of the free deformation, respectively. The influence of the bridge's free deformation (1) is widely distributed across the bridge body.
	\begin{figure}[!t]
		\centering
		\includegraphics[width=80mm]{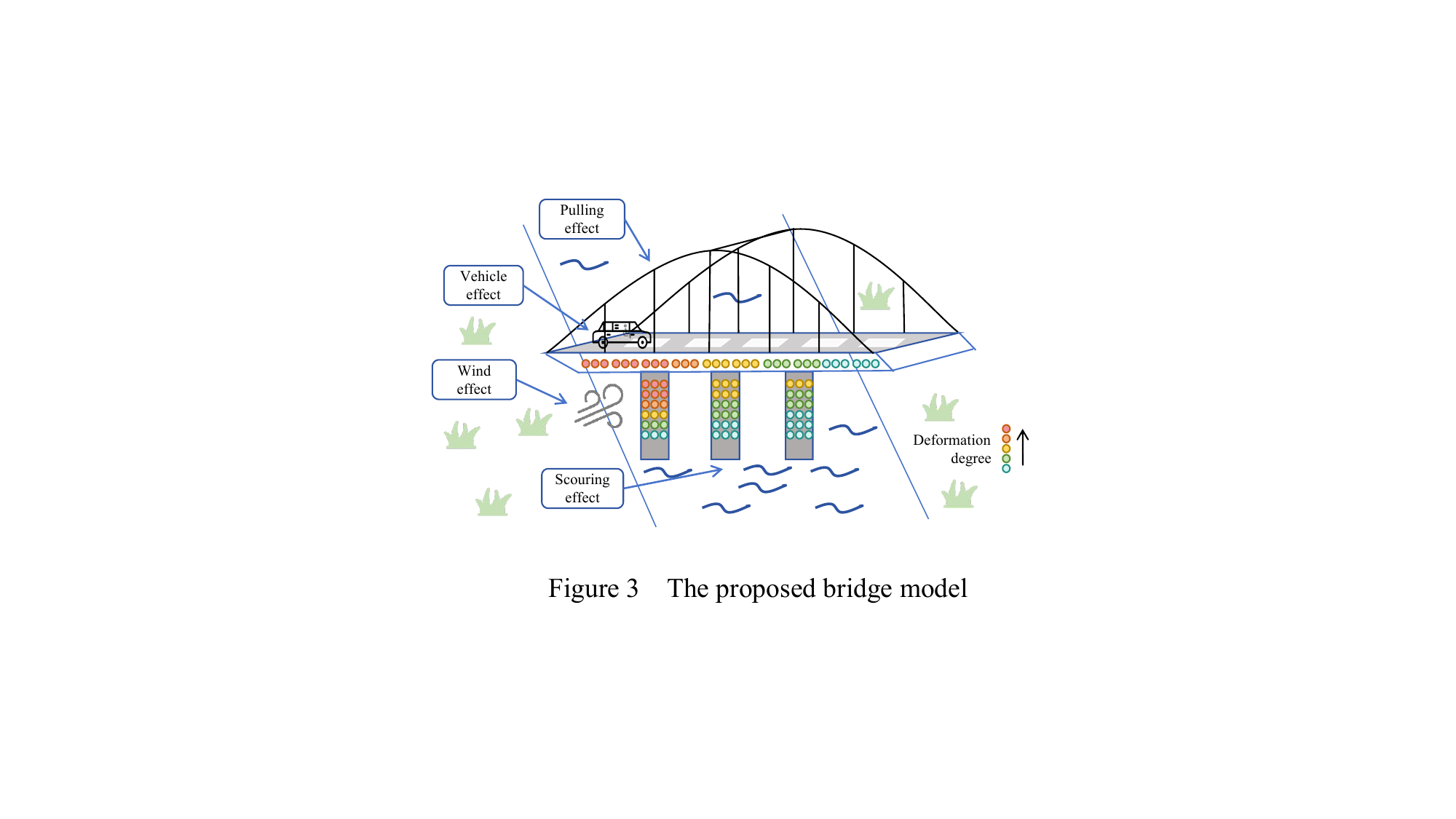}
		\caption{The forced deformation of the bridge.}
		\label{fig_2}
	\end{figure}
	
	While free deformation provides insights into the bridge's intrinsic behavior, it is equally critical to explore the bridge’s response to forced deformation.
	In the real physical world, various factors  can affect the vibration of bridges, such as vehicle-bridge and wind-bridge coupling effects, scouring effect from river currents, and pulling effect from the cables, as shown in Fig. 3. These effects are subsequently modeled as distinct excitation sources. 
	
	Assume that there are  $E$ excitation sources that affect the forced deformation of the bridge at different locations. Define \( L_{e} \) and \( v_{e} \) as initial position and velocity  of the \( i \)-th excitation source, where \( e = 1, 2, \ldots, E \), respectively. Moreover, denote \( \Delta L_e \) as the distance between the \( i \)-th excitation source and the BMDM point. The impact of the forced deformation on bridge can be expressed as
	\begin{equation}
		\begin{split}\color{black}
			\begin{aligned}
				D_{\text{forced}}=\sum_{e=1}^{E} A_{e} e^{-\zeta \Delta L_e(t)} \sin \left(2 \pi f_{e} t+\varphi_{B,e}\right),
			\end{aligned}
		\end{split}
	\end{equation}
	where $A_e$ is the amplitude of excitation from the $e$-th source, $f_e$ is the
	excitation frequency, and $\varphi_{B,e}$ denotes the initial phase of the $e$-th       
	excitation point. The exponential term $e^{-\zeta\Delta L_e(t)}$ accounts for the
	damping effect, where $\zeta$ is the damping coefficient, and the influence of an
	excitation source decreases as its distance $\Delta L_e(t)$ from the monitoring point  increases. This physically represents how external forces such as passing vehicles or wind loads  cause additional deformations that diminish with distance from the monitoring point.

	We propose to sample the deformation of the bridge once per frame. 
	Based on (1) and (2), the micro-deformation of the bridge at the $p$-th frame  can be expressed as
	\begin{equation}
		\begin{split}\color{black}
			\begin{aligned}
				\Delta D_{0\to p}  &=D_{\text{free}}+D_{\text{forced}} =A_{0} \sin \left(2 \pi f_{B} t+\varphi_{B}\right)\\
				&+\sum_{e=1}^{E} A_{e} e^{-\zeta \Delta L_e(t)} \sin \left(2 \pi f_{e} t+\varphi_{B,e}\right).
			\end{aligned}
		\end{split}
	\end{equation}
	
	Assume that the BS is positioned at the origin of the coordinate system, and the transmitting and receiving beamforming directions of the BS point directly toward the initial deformation position \(  (x^{\delta}_0, y^{\delta}_0, z^{\delta}_0) \), where \(\delta\) denotes parameters associated with bridge micro-deformation, as shown in Fig.~4. We perform the cartesian-to-spherical coordinate transformation whose formula is
	\begin{figure}[t!]
		\centering
		\includegraphics[width=85mm]{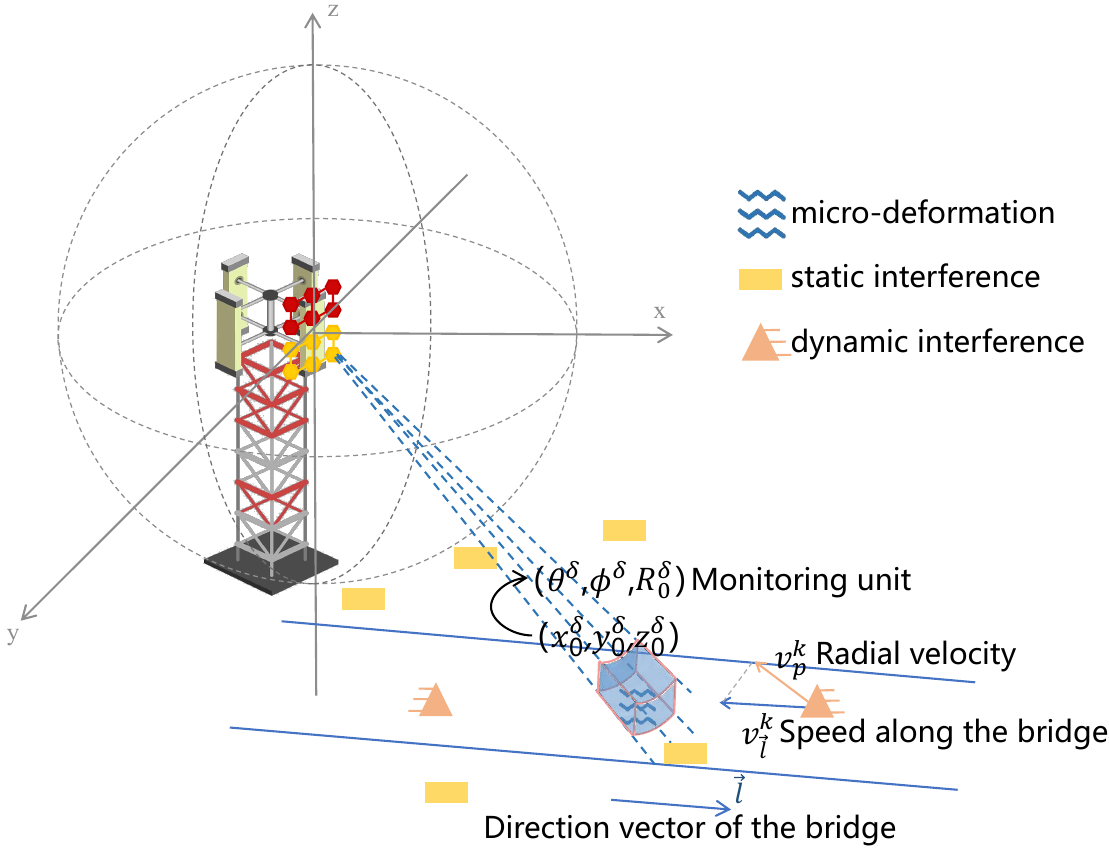}
		\caption{The diagram of the geometric transformation.}
		\label{fig_2}
	\end{figure}
	\begin{equation}
		\begin{split}\color{black}
			\begin{aligned}
				\begin{cases}
					\theta^{\delta} = \arctan\left(\frac{y^{\delta}_{0}}{x^{\delta}_{0}}\right), \\[1em]
					\phi^{\delta} = \arccos\left(\frac{z^{\delta}_{0}}{R^{\delta}_{0}}\right), \\[0.8em]
					R^{\delta}_{p} = \!\! \sqrt{\! \left(x^{\delta}_{0}\right)^2\!\!\! +\! \left(y^{\delta}_{0}\right)^2\!\!\! +\! \left(z^{\delta}_{0}\!\! +\!\! \Delta D_{0\to p}\right)^2},
				\end{cases}
			\end{aligned}
		\end{split}
	\end{equation}
	where $\theta ^{\delta},\phi ^{\delta}$, and $R^{\delta}_{p}$ represent the initial azimuthal angle, elevation angle, and the radial distance of the BMDM point  relative to the BS in the $p$-th frame, respectively. Moreover, $R^{\delta}_{p}$ is the core estimation objective of the BMDM problem, as it
	directly reflects the bridge micro-deformation $\Delta D_{0\to p}$. Then, the echo channel of micro-deformation on the $m$-th
	subcarrier of the $n$-th OFDM symbol in the $p$-th frame can be represented as
	\begin{equation}
		\begin{split}\color{black}
			\begin{aligned}
				\mathbf{H}_{p,n,m}^{\text{micro}} =\alpha ^{\delta} e^{-j\! \frac{4 \pi f_{m} R_{p}^{\delta}}{c}} \mathbf{a}_{R}\left(\Psi ^{\delta},\Omega ^{\delta}\right) \mathbf{a}_{H}^{T}\left(\Psi ^{\delta},\Omega ^{\delta}\right),
			\end{aligned}
		\end{split}
	\end{equation}
	where $\alpha ^{\delta}$ represents the micro-deformation echo channel fading factor,  $
	\Psi ^{\delta}=\Psi(\theta ^{\delta}, \phi ^{\delta}) = \cos \phi ^{\delta} \cos \theta ^{\delta}$ represents the horizontal spatial-domain direction, and 
	$
	\Omega ^{\delta}=\Omega(\theta ^{\delta}, \phi ^{\delta}) = \sin \phi ^{\delta} 
	$
	represents the pitch spatial-domain direction. Moreover, 
	$\mathbf{a}_R(\Psi ^{\delta}, \Omega ^{\delta})$ and $\mathbf{a}_H(\Psi ^{\delta}, \Omega ^{\delta})$ are the array steering vectors for the spatial-domain direction $(\Psi ^{\delta}, \Omega ^{\delta})$ of RU-UPA and HU-UPA with the form\cite{10854508}
	\begin{equation}
		\begin{split}\color{black}
			\begin{aligned}
				\!\mathbf{a}_{R}(\Psi ^{\delta}, \Omega ^{\delta}) =\mathbf{a}_{R}^{x}(\Psi ^{\delta}) \otimes \mathbf{a}_{R}^{z}(\Omega ^{\delta}) \in \mathbb{C}^{N_{R} \times 1} ,
			\end{aligned}
		\end{split}
	\end{equation}
	\begin{equation}
		\begin{split}\color{black}
			\begin{aligned}
				\mathbf{a}_{H}(\Psi ^{\delta}, \Omega ^{\delta}) =\mathbf{a}_{H}^{x}(\Psi ^{\delta}) \otimes \mathbf{a}_{H}^{z}(\Omega ^{\delta}) \in \mathbb{C}^{N_{H} \times 1},
			\end{aligned}
		\end{split}
	\end{equation}
	where $\otimes$ denotes the Kronecker product, and
	\begin{equation}
		\begin{split}\color{black}
			\begin{aligned}
				\!\!\!	\!\mathbf{a}_{R}^{x}(\Psi ^{\delta})\!\! =\!\! \left[1, e^{j \frac{2 \pi f_{c} d \Psi ^{\delta}}{c}}, \ldots, e^{j \frac{2 \pi f_{c} d \Psi ^{\delta}}{c}\left(N_{R}^{x}-1\right)}\right]^{T}\!\!\!\! \in \mathbb{C}^{N_{R}^{x} \times 1}, 
			\end{aligned}
		\end{split}
	\end{equation}
	\begin{equation}
		\begin{split}\color{black}
			\begin{aligned}
				\!\!\!\!\!	\mathbf{a}_{R}^{z}(\Omega ^{\delta})\!\! =\!\! \left[1, e^{j \frac{2 \pi f_{c} d \Omega ^{\delta}}{c}}, \ldots, e^{j \frac{2 \pi f_{c} d \Omega ^{\delta}}{c}\left(N_{R}^{z}-1\right)}\right]^{T}\!\!\!\! \in \mathbb{C}^{N_{R}^{z} \times 1}, 	
			\end{aligned}
		\end{split}
	\end{equation}
	\begin{equation}
		\begin{split}\color{black}
			\begin{aligned}
				\!\!\!\!	\!\mathbf{a}_{H}^{x}(\Psi ^{\delta})\!\! =\!\! \left[\!1, e^{j \frac{2 \pi f_{c} d \Psi ^{\delta}}{c}}, \ldots, e^{j \frac{2 \pi f_{c} d \Psi ^{\delta}}{c}\left(N_{H}^{x}-1\right)} \!\right]^{T}\!\!\!\! \!\! \in\!  \mathbb{C}^{N_{H}^{x} \times 1}, 
			\end{aligned}
		\end{split}
	\end{equation}
	\begin{equation}
		\begin{split}\color{black}
			\begin{aligned}
				\!\!\!\!\!	\mathbf{a}_{H}^{z}(\Omega ^{\delta})\!\! =\!\! \left[\!1, e^{j \frac{2 \pi f_{c} d \Omega ^{\delta}}{c}}, \ldots, e^{j \frac{2 \pi f_{c} d \Omega ^{\delta}}{c}\left(N_{H}^{z}-1\right)} \!\right]^{T}\! \!\!\!\!\!  \in\!  \mathbb{C}^{N_{H}^{z} \times 1} .
			\end{aligned}
		\end{split}
	\end{equation}
	\subsection{Echo Channel Model of Interference}
	As shown in Fig.~4, besides dynamic interferences such as vehicles and pedestrians passing through the bridge, there are also numerous static interferences within the scanning range of the BS.
	
	Assume that there are $K$ dynamic interferences that move along the direction vector $\vec{l}=(l_{x},l_{y},l_{z})$ of the bridge, and the velocity of the \( k \)-th dynamic interference along this direction is denoted as \( v_{\vec{l}}^k \in \mathbb{R} \), \( k = 1, 2, \ldots, K \). These dynamic interferences may not only affect the micro-deformation of the bridge but also impact the echo signals. To obtain the  echo channel of the dynamic interferences, we first transform  $v_{\vec{l}}^{k}$ into the BS-relative radial velocity $v_{p}^{k} \in \mathbb{R}$ in the $p$-th frame, and the transformation  can be represented as 
	\begin{equation}
		\begin{split}\color{black}
			\begin{aligned}
				\begin{cases}
					\mathbf{\vec{v}}_{p}^{k}  & = \frac{\vec{l} }{\|\vec{l} \|} \cdot v_{\vec{l}}^{k},\\[0.8em]
					L_{p}^{k}&=L_{0}^{k}+\mathbf{\vec{v}}_{p}^{k} \cdot pT_{f},\\[0.8em]
					R_{p}^{k}&=\|L_{p}^{k} \|,\\[0.8em]
					v_{p}^{k}&=\sum (\mathbf{\vec{v}}_{p}^{k} \odot \frac{L_{p}^{k}}{R_{p}^{k}} ),
				\end{cases}
			\end{aligned}
		\end{split}
	\end{equation}
	where $\odot$ represents element wise multiplication,    while  $\mathbf{\vec{v}}_{p}^{k}  \in \mathbb{R}^3$, $L_p^k \in \mathbb{R}^3$ and $R_p^k \in \mathbb{R}^+$ represent the 3D velocity vector along direction $\vec{l}$, the 3D spatial position, and the radial distance to the BS of the $k$-th dynamic interference in the $p$-th frame, respectively. Then, the echo channel of dynamic interferences on the $m$-th
	subcarrier of the $n$-th OFDM symbol in the $p$-th frame can be represented as
	\begin{equation}
		\begin{split}\color{black}
			\begin{aligned}
				\mathbf{H}_{p,n,m}^{\text{dynamic}} =\sum_{k=1}^{K} \alpha ^{k }_{p}&e^{-j  \frac{4 \pi f_{m} R_{p}^{k}}{c}} e^{j 4 \pi f_{c} \frac{ v_{p}^{k} n T_{\text {sym}}}{c}}\times \\
				&\,\,\,\,\,\,\,\,\,\,\,\,\,\,\,\,\,\,\, \mathbf{a}_{R}\left(\Psi ^{\delta},\Omega ^{\delta}\right) \mathbf{a}_{H}^{T}\left(\Psi ^{\delta},\Omega ^{\delta}\right),
			\end{aligned}
		\end{split}
	\end{equation}
	where  $\alpha ^{k }_{p}$ denotes the channel fading factor of the $k$-th dynamic interference in the $p$-th frame.
	
	Meanwhile,
	assume that there are $S$ static interferences within the scanning area of the BS. Then, the echo channel of static interferences on the $m$-th
	subcarrier of the $n$-th OFDM symbol in the $p$-th frame can be represented as\cite{10477890}
	\begin{equation}
		\begin{split}\color{black}
			\begin{aligned}
				\mathbf{H}_{p,n,m}^{\text{static}} =\sum_{s=1}^{S} \alpha ^{s}e^{-j  \frac{4 \pi f_{m} R^{s}}{c}}
				\mathbf{a}_{R}\!\left(\Psi^{\delta},\!\Omega^{\delta}\right)\! \mathbf{a}_{H}^{T}\!\!\left(\Psi^{\delta},\!\Omega^{\delta}\right),
			\end{aligned}
		\end{split}
	\end{equation}
	where  $\alpha ^{s}$ and $R^{s}$ denote the channel fading factor and radial distance to BS of the $s$-th static interference, respectively.
	
	\subsection{Echo Signals of Micro-Deformation}
	Assume that the BS performs beamforming with the array
	steering vector $\mathbf{a}_{H} (\Psi^{\delta}, \Omega^{\delta})$ through HU-UPA. The transmission signal on the \( m \)-th subcarrier of the \( n \)-th OFDM symbol in the \( p \)-th frame is 
	
	\begin{equation}
		\mathbf{x}_{p,n,m} = \sqrt{\frac{\rho_s P_t}{N_{H}}} \mathbf{a}_{H} (\Psi^{\delta}, \Omega^{\delta}) q_{p,n,m}, 
	\end{equation}
	where \( P_t \) denotes the transmission power of BS, \( \rho_s \) denotes the power allocation factor towards the initial position of the BMDM point $  (x^d_0, y^d_0, z^d_0) $, and \( q_{p,n,m} \) denotes the BMDM signal\cite{9945983}. Meanwhile, we assume that RU-UPA of the BS also performs receive beamforming with 
	$\mathbf{w} = \frac{1}{\sqrt{N_R}} \mathbf{a}_{R} (\Psi^{\delta},\Omega^{\delta}).$
	Then, the frequency-domain echo signal received by BS on the \( m \)-th subcarrier of the \( n \)-th OFDM symbol in the \( p \)-th frame can be represented as 
	\begin{equation}
		\begin{split}\color{black}
			\begin{aligned}
				&y_{p,n,m} = \mathbf{w}^H \mathbf{H}_{p,n,m}^{\text{channel}}\mathbf{x}_{p,n,m}^{\ast} + n_{p,n,m}\\[0.8em]
				&=\mathbf{w}^H (\mathbf{H}_{p,n,m}^{\text{micro}}+\mathbf{H}_{p,n,m}^{\text{dynamic}}+\mathbf{H}_{p,n,m}^{\text{static}})\mathbf{x}_{p,n,m}^{\ast} + n_{p,n,m},
			\end{aligned}
		\end{split}
	\end{equation}
	where \( \mathbf{H}_{p,n,m}^{\text{channel}} = \mathbf{H}_{p,n,m}^{\text{micro}}+\mathbf{H}_{p,n,m}^{\text{dynamic}}+\mathbf{H}_{p,n,m}^{\text{static}} \). Moreover, \( n_{p,n,m} \) is zero-mean additive Gaussian noise with variance \( \sigma^2 \). After receiving the echo signal \( y_{p,n,m} \), 
	we could remove the effect of \( q_{p,n,m} \) from
	\begin{equation}
		\check{y}_{p,n,m} = \frac{y_{p,n,m}}{q_{p,n,m}}.
	\end{equation}
	
	Then, we concatenate the echo signals on all subcarriers, all OFDM symbols and all frames into a tensor \( \mathbf{Y} \in \mathbb{C}^{P \times N \times M} \), with \( \mathbf{Y}[p,n,m] = \check{y}_{p,n,m} \), \( p = 0, \ldots, P - 1 \), \( n = 0, \ldots, N - 1 \), and \( m = 0, \ldots, M - 1 \).

	\section{Interference Suppression Scheme}
	In this section,  we utilize phasor statistical analysis to suppress the dynamic clutter and static clutter on BMDM. 
	
	\newcounter{TempEqCnt18} % 创建临时变量TempEqCnt
	\setcounter{TempEqCnt18}{\value{equation}} % 将当前公式序号 赋给TempEqCnt
	\setcounter{equation}{17} % 当前公式序号变为x，x等于长公式应有的序号减1.
	\begin{figure*}[ht] %hb代表放在文章底部，%ht为放在文章顶部
		\begin{equation}
			\begin{split}\color{black}
				\begin{aligned}
					\mathbf{Y}[p,n,m] &= \check{y}_{p,n,m} = \mathbf{w}^H (\mathbf{H}_{p,n,m}^{\text{micro}}+\mathbf{H}_{p,n,m}^{\text{dynamic}}+\mathbf{H}_{p,n,m}^{\text{static}})\mathbf{\check{x}}^{*} + \check{n}_{p,n,m}, \\
					&=\underbrace{G^{\delta}e^{-j \frac{4 \pi f_{m} R_{p}^{\delta}}{c}} }_{\mathrm{micro-deformation}}+
					\underbrace{\textstyle \sum_{k=1}^{K} G^{k}_{p}e^{-j  \frac{4 \pi f_{m} R_{p}^{k}}{c}} e^{j 4 \pi f_{c} \frac{ v_{p}^{k} n T_{\text {sym}}}{c}}}_{\mathrm{dynamic~interference}}
					+\underbrace{\textstyle \sum_{s=1}^{S} G^{s}e^{-j  \frac{4 \pi f_{m} R^{s}}{c}}}_{\mathrm{static~interference}}
					+\underbrace{\check{n}_{p,n,m}}_{\mathrm{Noise}}.
				\end{aligned}
			\end{split}
		\end{equation}
		
		\vspace*{-5mm} % 调整线与公式之间的距离
	\end{figure*}
	\newcounter{TempEqCnt22} % 创建临时变量TempEqCnt
	\setcounter{TempEqCnt22}{\value{equation}} % 将当前公式序号 赋给TempEqCnt
	\setcounter{equation}{21} % 当前公式序号变为x，x等于长公式应有的序号减1.
	\begin{figure*}[ht] %hb代表放在文章底部，%ht为放在文章顶部
		\begin{equation}
			\begin{split}\color{black}
				\begin{aligned}
					\mathbf{F}[p,n,\kappa^{\delta}]
					=&\frac{G^{\delta}}{M} e^{j 2 \pi\big(-f_{c} \frac{2 R_{p}^{\delta}}{c}+\big(-\Delta f \frac{2 R_{p}^{\delta}}{c}+\frac{\kappa^{\delta}}{M}\big) \frac{M-1}{2}\big)}\frac{\sin \big[\big(-\pi\Delta f \frac{2R_{p}^{\delta}}{c} + \frac{\pi}{M}\kappa^{\delta}\big) M\big]}{\sin \big[\big(-\pi\Delta f \frac{2R_{p}^{\delta}}{c} + \frac{\pi}{M}\kappa^{\delta}\big)\big]} \\
					&\!\!+\frac{1}{M} \sum_{k=1}^{K} G_{p}^{k} e^{j 2 \pi\big(-f_{c} \frac{2 R_{p}^{k}}{c}+\big(-\Delta f \frac{2 R_{p}^{k}}{c}+\frac{\kappa^{\delta}}{M}\big) \frac{M-1}{2}+f_{c} \frac{2 v_{p}^{k} nT_{\text {sym}}}{c}\big)} \frac{\sin \big[\big(-\pi\Delta f \frac{2R_{p}^{k}}{c} + \frac{\pi}{M}\kappa^{\delta}\big) M\big]}{\sin \big[\big(-\pi\Delta f \frac{2R_{p}^{k}}{c} + \frac{\pi}{M}\kappa^{\delta}\big)\big]} \\
					&\!\!+\frac{1}{M} \sum_{s=1}^{S} G^{s} e^{j 2 \pi\big(-f_{c} \frac{2 R^{s}}{c}+\big(-\Delta f\frac{2R^{s}}{c} +\frac{\kappa^{\delta}}{M}\big) \frac{M-1}{2}\big)}\frac{\sin \big[\big(-\pi\Delta f \frac{2R^{s}}{c} + \frac{\pi}{M}\kappa^{\delta}\big) M\big]}{\sin \big[\big(-\pi\Delta f \frac{2R^{s}}{c} + \frac{\pi}{M}\kappa^{\delta}\big)\big]} 
					+\check{n}_{p, n}^{\prime}.
				\end{aligned}
			\end{split}
		\end{equation}
		
		\vspace*{1mm} % 调整线与公式之间的距离
		\hrulefill % 添加一条水平线
	\end{figure*}

	\begin{figure}[!t]
		\centering
		\includegraphics[width=85mm]{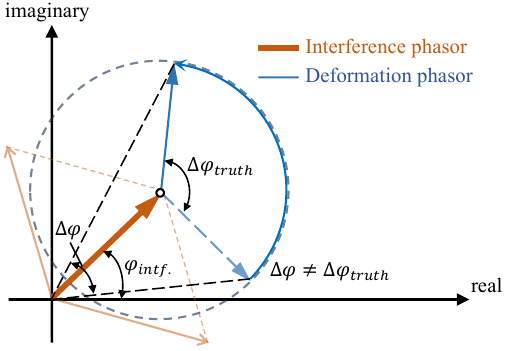}
		\caption{The diagram of the interference effect.}
		\label{fig_1}
	\end{figure}
	\subsection{Dynamic Interference Clutter Suppression}
	As shown in Fig.~5, the deformation phasor causes a phase change in the echo signal, while the interference phasor alters both the amplitude and phase of the echo signal, which leads to measurement error.
	
	Substituting (5), (13), and (14) into (16) and (17), we can reformulate the echo signal into (18), as shown on the top of this page, where $\mathbf{\check{x}}^{*} = \frac{\mathbf{x}_{p,n,m}^{*}}{q_{p,n,m}}$ and $\check{n}_{p,n,m} = \frac{n_{p,n,m}}{q_{p,n,m}}$. Moreover, $G^{\delta}$, $G^{k}_{p}$, and $G^{s}$ in (18) are defined as
	\newcounter{TempEqCnt19} % 创建临时变量TempEqCnt
	\setcounter{TempEqCnt19}{\value{equation}} % 将当前公式序号 赋给TempEqCnt
	\setcounter{equation}{18} % 当前公式序号变为x，x等于长公式应有的序号减1.
	\begin{equation}
		G^{\delta}  = \alpha ^{\delta} \mathbf{w}^{H}  \mathbf{a}_{R}\left(\Psi^{\delta}, \Omega^{\delta}\right)  \mathbf{a}_{H}^{T}\left(\Psi ^d, \Omega^{\delta}\right) \mathbf{\check{x}}^{*}, 
	\end{equation}
	\begin{equation}
		G_{p}^{k} = \alpha ^{k }_{p} \mathbf{w}^{H}   \mathbf{a}_{R}\left(\Psi^{\delta}, \Omega^{\delta}\right)  \mathbf{a}_{H}^{T}\left(\Psi ^d, \Omega^{\delta}\right) \mathbf{\check{x}}^{*}, 
	\end{equation}
	\begin{equation}
		G^{s} = \alpha ^{s} \mathbf{w}^{H}   \mathbf{a}_{R}\left(\Psi^{\delta}, \Omega^{\delta}\right)  \mathbf{a}_{H}^{T}\left(\Psi ^d, \Omega^{\delta}\right) \mathbf{\check{x}}^{*}.
	\end{equation}

	Eq.~(18) shows the four components of the echo signal: the effective  micro-deformation  echo, dynamic interference clutter, static interference clutter, and the noise. Next, we suppress the dynamic interferences and the static interferences in the echo signal $\mathbf{Y}[p,n,m]$ to obtain the effective micro-deformation echo.
	
	Let us apply an $M$-point IDFT on $\mathbf{Y}[p,n,:]$ along the subcarrier dimension to  preliminarily separate BMDM points and various interferences, and obtain $\mathbf{F}[p,n,\kappa]\!\! =\!\! \text{IDFT}(\mathbf{Y}[p,n,:])$, $\kappa = 0, \ldots, M - 1$. 
	
	Note that $\mathbf{F}[p,n,\kappa^{\delta}]$ contains the micro-deformation information that can be extracted from $\mathbf{F}[p,n,\kappa]$, as shown in (22) on the top of this page, where $\kappa^{\delta}=\left \lfloor  \frac{M\Delta f 2 R_{0}^{\delta}}{c} \right \rceil $ represents the range bin index that contains the micro-deformation point\cite{5776640}. However, there is still energy leakage between different range bins after IDFT. Due to the high accuracy requirements of BMDM, we should further remove the remaining interference in $\mathbf{F}[p,n,\kappa^{\delta}]$, i.e., the second and the third terms in (22).

	%\subsubsection{Removal of Single Dynamic Interference}

	As shown in Fig. 6(a) and Fig. 6(c), the impact of single dynamic interference differs from that of multiple dynamic interferences on BMDM. This distinction can be observed through the amplitude and phase
	variation of the second term in (22). To eliminate the dynamic clutter in different situations, we first design a circle-fitting-based single dynamic interference removal (CF-SDIR) method. For clarity, we redefine the constant in (19) as
	\newcounter{TempEqCnt23} % 创建临时变量TempEqCnt
	\setcounter{TempEqCnt23}{\value{equation}} % 将当前公式序号 赋给TempEqCnt
	\setcounter{equation}{22} % 当前公式序号变为x，x等于长公式应有的序号减1.
	\begin{equation}
		\begin{split}\color{black}
			\begin{aligned}
				Z_{p}^{\delta}=G^{\delta}e^{j\pi \frac{\kappa^{\delta}(M-1)}{M} } {\scriptsize  \frac{\sin [(-\pi\Delta f \frac{2R_{p}^{\delta}}{c} + \frac{\pi}{M}\kappa^{\delta}) M]}{\sin [(-\pi\Delta f \frac{2R_{p}^{\delta}}{c} + \frac{\pi}{M}\kappa^{\delta})]}},	
			\end{aligned}
		\end{split}
	\end{equation}
	\begin{figure*}[!t]
		\centering
		\subfloat[]{\includegraphics[width=75mm]{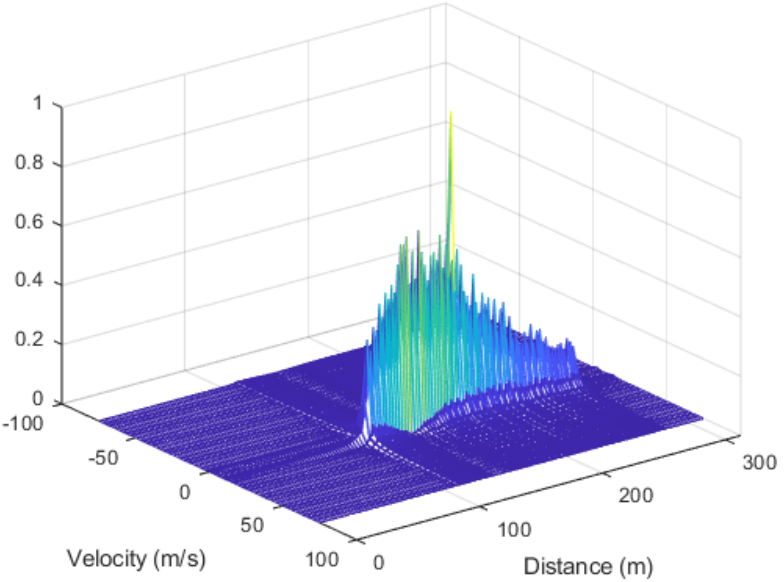}%
			\label{fig_first_case}}
		\hfil
		\subfloat[]{\includegraphics[width=75mm]{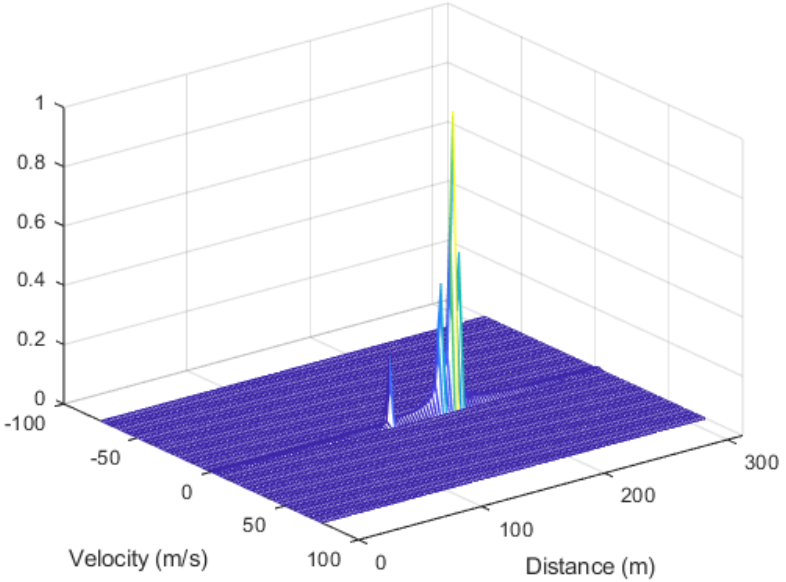}%
			\label{fig_second_case}}
		\hfil
		\subfloat[]{\includegraphics[width=75mm]{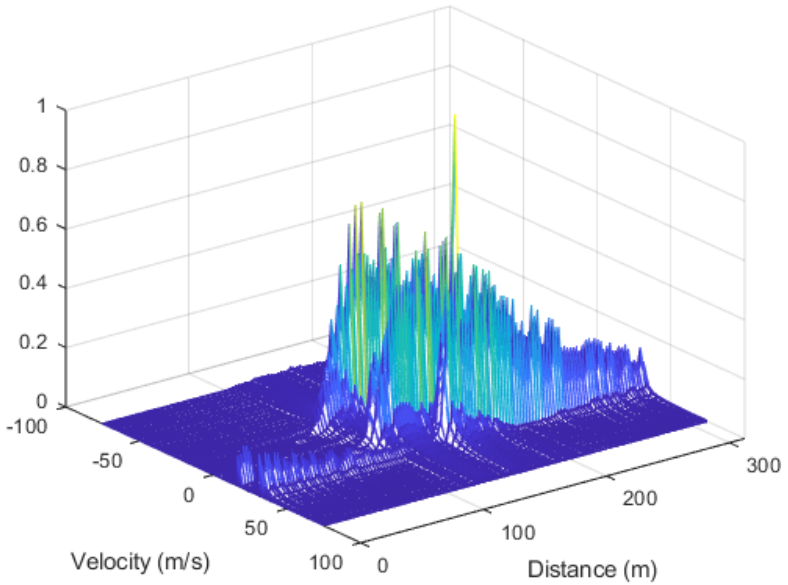}%
			\label{fig_first_case}}
		\hfil
		\subfloat[]{\includegraphics[width=75mm]{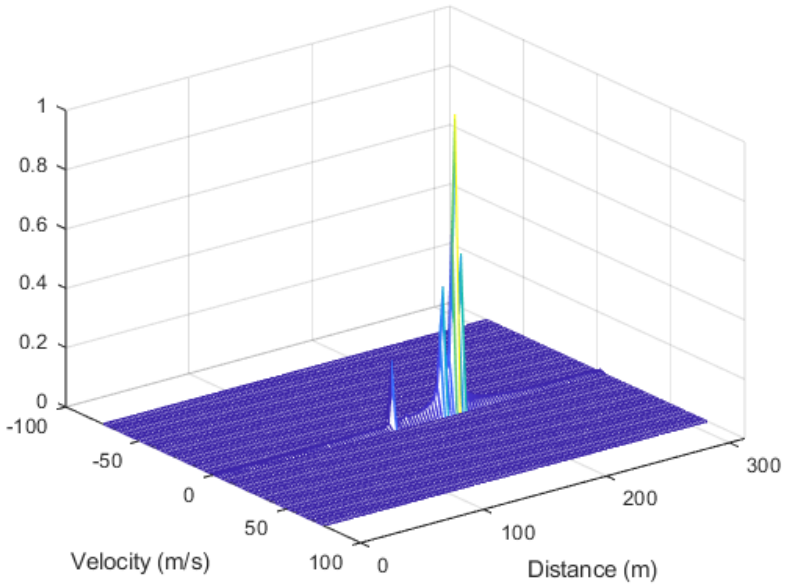}%
			\label{fig_second_case}}
		\caption{(a) Range-Doppler spectrum of multi-frame echo signals with single dynamic interference. 
			(b)  Range-Doppler spectrum of multi-frame echo signals with CF-SDIR. 
			(c) Range-Doppler spectrum of multi-frame echo signals with $3$ dynamic interferences.
			(d) Range-Doppler spectrum of multi-frame echo signals with PM-MDIS.
			\textcolor{black}{The monitoring point is set as $(180,60,-25)$ m.}}
		\label{fig_sim}
	\end{figure*}
	\begin{equation}
		\begin{split}\color{black}
			\begin{aligned}
				Z_{p}^{k}=G^{k}_{p}e^{j\pi \frac{\kappa^{\delta}(M-1)}{M} } {\scriptsize  \frac{\sin [(-\pi\Delta f \frac{2R_{p}^{k}}{c} + \frac{\pi}{M}\kappa^{\delta}) M]}{\sin [(-\pi\Delta f \frac{2R_{p}^{k}}{c} + \frac{\pi}{M}\kappa^{\delta})]}},	
			\end{aligned}
		\end{split}
	\end{equation}
	\begin{equation}
		\begin{split}\color{black}
			\begin{aligned}
				Z^{s}=G^{s}e^{j\pi \frac{\kappa^{\delta}(M-1)}{M} } {\scriptsize  \frac{\sin [(-\pi\Delta f \frac{2R^{s}}{c} + \frac{\pi}{M}\kappa^{\delta}) M]}{\sin [(-\pi\Delta f \frac{2R^{s}}{c} + \frac{\pi}{M}\kappa^{\delta})]}},
			\end{aligned}
		\end{split}
	\end{equation}
	where $Z_{p}^{\delta}\approx Z^{\delta}=MG^{\delta}e^{j\pi \frac{\kappa^{\delta}(M-1)}{M}}$ due to $\kappa^{\delta}=\left \lfloor  \frac{M\Delta f 2 R_{p}^{\delta}}{c} \right \rceil  $. 
	Consequently, $\mathbf{F}[p,n,\kappa^{\delta}]$ can be simplified as 
	\begin{equation}
		\begin{split}\color{black}
			\begin{aligned}
				\mathbf{F}[p,n,\kappa^{\delta}] = &Z^{\delta}e^{j2\pi (-\frac{2f_{c}+\Delta f(M-1)}{c} )R_{p}^{\delta}}\\
				&\!\!\!\!\!\!+{ \sum_{s=1}^{S}} Z^{s}e^{j2\pi (-\frac{2f_{c}+\Delta f(M-1)}{c})R^{s}} +\check{n}_{p, n}^{\prime} \\
				&\!\!\!\!\!\!+Z_{p}^{k^{*}}e^{j2\pi (-\frac{2f_{c}+\Delta f(M-1)}{c} )R_{p}^{k^{*}}}e^{j2\pi f_{p,\text{dynamic} }^{k^{*}}n},
			\end{aligned}
		\end{split}
		\tag{26}
	\end{equation}
	where $f_{p,\text{dynamic} }^{k^{*}} = \frac{2f_{c}v_{p}^{k^{*}}T_{\text {sym}}}{c}$. 	Moreover, $k^{*}$ represents that only the $k^{*}$-th dynamic target exists in the vicinity of the monitoring point.
	The first two terms in (26) that represent the micro-deformation and static interference components of the echo after IDFT processing remain unchanged across different OFDM symbols in the $p$-th frame. The last term in (26) represents the single dynamic interference component of the echo after IDFT processing, whose phase varies linearly with respect to $n$. Therefore, in case noise is ignored, $\mathbf{F}[p,n,\kappa^{\delta}]$ can be further rewritten as
	\begin{equation}
		\begin{split}\color{black}
			\begin{aligned}
				\mathbf{F}[p,n,\kappa^{\delta}]=O_{p,\text{dynamic}}+U_{\text{dynamic}}e^{j2\pi f^{k}_{p,\text{dynamic}}n}  ,
			\end{aligned}
		\end{split}
	\end{equation}
	where 
	\begin{equation}
		\begin{split}\color{black}
			\begin{aligned}
				O_{p,\text{dynamic}} =\, &Z^{\delta} e^{j2\pi (-\frac{2f_{c}+\Delta f(M-1)}{c} )R_{p}^{\delta}}
				\\&\,\,  + { \textstyle\sum_{s=1}^{S}} Z^{s}e^{j2\pi (-\frac{2f_{c}+\Delta f(M-1)}{c})R^{s}},\\[0.5em]
			\end{aligned}
		\end{split}
	\end{equation}
	\vspace*{-2mm} % 调整线与公式之间的距离
	\begin{equation}
		\begin{split}\color{black}
			\begin{aligned}
				\!\!\!\!\!\!\!\!\!\!\!\!\!\!\!\!\!\!\!\!\!\!\!\!\!\!\!\!	U_{\text{dynamic}}=Z_{p}^{k}e^{j2\pi (-\frac{2f_{c}+\Delta f(M-1)}{c} )R_{p}^{k}}.
			\end{aligned}
		\end{split}
	\end{equation}
	
	Subsequently, we employ the least squares method to perform an $N$-point circle fitting on $\mathbf{F}[p,:,\kappa^{\delta}]$, and obtain the circle center $O_{p,\text{dynamic}}$, which effectively removes the single dynamic interference clutter component from $\mathbf{F}[p,n,\kappa^{\delta}]$, as shown in Fig.~6(b). Then, the complex scalar circle center $O_{p,\text{dynamic}}$ is stored in the vector $\mathbf{O}[p]$  to remove the static interference clutter.
	
	The CF-SDIR method is only applicable when there is a single dynamic interference on the bridge. However, since we usually do not know how many dynamic interferences are on the bridge, we directly perform circle-fitting on $\mathbf{F}[p,:,\kappa^{\delta}]$. If $\mathbf{F}[p,:,\kappa^{\delta}]$ can be fitted into a circle, then there is only one dynamic interference, and we could utilize CF-SDIR. Otherwise, there are  multiple dynamic interferences on the bridge, and the expression of $\mathbf{F}[p,n,\kappa^{\delta}]$ could be formulated as 
	\begin{equation}
		\begin{split}\color{black}
			\begin{aligned}
				\mathbf{F}[p,n,\kappa^{\delta}] = &Z^{\delta}e^{j2\pi (-\frac{2f_{c}+\Delta f(M-1)}{c} )R_{p}^{\delta}}\\
				&+{ \sum_{s=1}^{S}} Z^{s}e^{j2\pi (-\frac{2f_{c}+\Delta f(M-1)}{c})R^{s}} +\check{n}_{p, n}^{\prime} \\
				&+{ \sum_{k=1}^{K}}Z_{p}^{k}e^{j2\pi (-\frac{2f_{c}+\Delta f(M-1)}{c} )R_{p}^{k}}e^{j2\pi f_{p }^{k}n},
			\end{aligned}
		\end{split}
	\end{equation}
	where $f_{p }^{k} = \frac{2f_{c}v_{p}^{k}T_{\text {sym}}}{c}$. 
	In the $p$-th frame, due to the parameters \( Z_p^k \), \( R_p^k \), and \( f_p^k \) vary across each distinct dynamic interference, $\mathbf{F}[p,:,\kappa^{\delta}]$ can no longer fit into a circle. In this case, we could design the phasor-mean-based multiple dynamic interferences suppression (PM-MDIS) method to suppress the  dynamic clutter. Specifically, we note that the last term in (30) which represents the dynamic clutter can be suppressed by the following formula as
	\begin{equation}
		\begin{split}\color{black}
			\begin{aligned}
				\lim_{N \to \infty}\left[\frac{1}{N}\sum_{n=0}^{N-1}\sum_{k=1}^{K}Z_{p}^{k}e^{j2\pi (-\frac{2f_{c}+\Delta f(M-1)}{c} )R_{p}^{k}}e^{j2\pi f_{p }^{k}n}   \right]=0. 
			\end{aligned}
		\end{split}
	\end{equation}
	
	As shown in (31), the phase of $e^{j2\pi f_{p}^{k}n}$ for each dynamic interference varies with $n$, which causes their corresponding vectors to undergo periodic rotation.\begin{figure*}[!t]
		\centering
		\subfloat[]{\includegraphics[width=85mm]{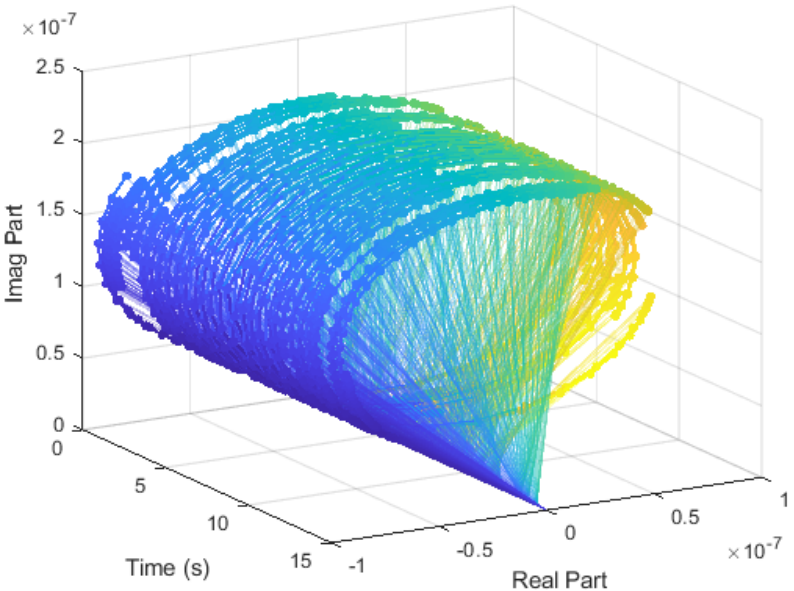}%
			\label{fig_first_case}}
		\hfil
		\subfloat[]{\includegraphics[width=85mm]{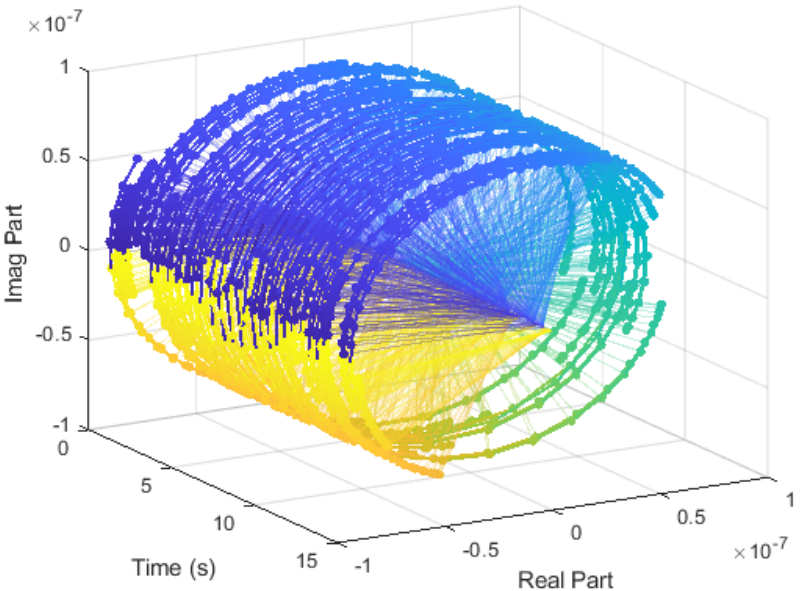}%
			\label{fig_second_case}}
		\caption{
			(a) Amplitude-phase diagram of complex number in multiple frames with dynamic interference suppression.
			(b) Amplitude-phase diagram of complex number in multiple frames with dynamic interference suppression and CF-MSIR.
			\textcolor{black}{The monitoring point is set as $(180,60,-25)$ m.}}
		\label{fig_sim}
	\end{figure*}

 By taking the average of $\mbox{$\mathbf{F}[p,:,\kappa^{\delta}]$}$, we could obtain a complex scalar $o_{p}$ that excludes multiple dynamic interferences as
	\begin{equation}
		\begin{split}\color{black}
			\begin{aligned}
				o_{p}=&\frac{1}{N}\sum_{n=0}^{N-1}\mathbf{F}[p,n,\kappa^{\delta}]= Z^{\delta}e^{j2\pi (-\frac{2f_{c}+\Delta f(M-1)}{c} )R_{p}^{\delta}}
				\\&+{ \sum_{s=1}^{S}} Z^{s}e^{j2\pi (-\frac{2f_{c}+\Delta f(M-1)}{c})R^{s}}+\frac{1}{N}\sum_{n=0}^{N-1}\check{n}_{p, n}^{\prime} .
			\end{aligned}
		\end{split}
	\end{equation}
	Furthermore, due to $\lim_{N \to \infty}\left[\frac{1}{N}\sum_{n=0}^{N-1}\check{n}_{p, n}^{\prime}\right]=0$, the PM-MDIS method also demonstrates excellent suppression of Gaussian white noise in the echo signal. By storing the complex scalar $o_{p}$ into vector $\mathbf{O}[p]$, the multiple dynamic interferences clutter in $\mathbf{F}[p,n,\kappa^{\delta}]$ could be suppressed, as shown in Fig. 6(d). Then, the vector $\mathbf{O}[p]$  can serve as the input for the subsequent approach to remove static interfence clutter.
	
		Unfortunately, in practical systems, $N$ can not approach infinity. 
		Let us define $\epsilon_{p}^{\text{dynamic}}$ as the dynamic interference
		margin caused by finite $N$.
		When N is sufficiently large to satisfy the velocity resolution requirement  
		$\frac{c}{2f_0T_{\text{sym}}N} < 1$ m/s, we can first perform DFT on OFDM
		symbol dimension of $\mathbf{F}[p,n,\kappa^{\delta}]$ to preliminarily estimate each
		$v_{p}^{k}$. Then we calculate the number of symbols required for each dynamic
		interference vector to complete a $2\pi$ phase rotation using $N_{p}^{k} =
		\frac{c}{2f_0v_{p}^{k}T_s}$. Finally, we select their least common multiple as the
		number of symbols utilized by PM-MDIS, which achieves $\epsilon_{p}^{\text{dynamic}} = 0$.      
		Let us define this process as the improved PM-MDIS (IPM) method.
	
		When $N$ is small, say when only tens of symbols can be used for sensing to ensure      
		communication efficiency, the effectiveness of PM-MDIS is guaranteed by $v_{p}^{k} =        
		\gamma_{p}^{k} \cdot \frac{c}{2f_0NT_s}$, where $\gamma_{p}^{k} \in \mathbb{Z}$ is an       
		integer. When each dynamic interference velocity $v_{p}^{k}$ can find a corresponding       
		$\gamma_{p}^{k}$ that satisfies  $v_{p}^{k} =        
		\gamma_{p}^{k} \cdot \frac{c}{2f_0NT_s}$ to make $\epsilon_{p}^{\text{dynamic}} = 0$, the PM-MDIS can completely filter out the dynamic interference. When there exists
		$v_{p}^{k}$ that can not find its corresponding $\gamma_{p}^{k}$ to satisfies that satisfies  $v_{p}^{k} =        
		\gamma_{p}^{k} \cdot \frac{c}{2f_0NT_s}$, there is  
		$\epsilon_{p}^{\text{dynamic}} \neq 0$, and PM-MDIS can only suppress the dynamic
		interference rather than completely filtering it out. In this case, the effectiveness of    
		PM-MDIS is related to the distance between $\frac{2f_0v_{p}^{k}NT_s}{c}$ and the
		nearest integer, where the closer the distance is to $0$, the more effective it becomes.    
		CF-SDIR can improve the interference suppression performance when only one dynamic
		interference exists in this situation.

	For clarity, we describe a specific operating mechanism for dynamic interference       
	suppression that jointly employs CF-SDIR or PM-MDIS: Since the number of dynamic interferences on the bridge deck is unknown, after obtaining    
	the feature vector $\mathbf{F}[p,n,\kappa^{\delta}]$, we first assume its distribution     
	is circular and perform fitting. For each data point, we calculate the difference
	between its distance to the fitted circle center and the fitted radius. An adaptive
	threshold is set to filter qualified points, and the proportion of qualified points is      
	utilized to determine whether the distribution is circular. If it is circular, then CF-SDIR      
	is applied. Otherwise, PM-MDIS is adopted. This process enables the suppression of      
	dynamic interference, which is defined as the CPM method.

	\subsection{Static Interference Clutter Suppression}
	
	We develop a circle-fitting-based multiple static interferences removal (CF-MSIR) method to eliminate the static clutter in $\mathbf{O}[p]$.
	
	In case the noise $\check{n}_{p, n}^{\prime}$ are ignored, $\mathbf{O}[p]$ could be expressed as
	\begin{equation}
		\begin{split}\color{black}
			\begin{aligned}
				\mathbf{O}[p]& = Z^{\delta}e^{j2\pi (-\frac{2f_{c}+\Delta f(M-1)}{c} )R_{p}^{\delta}}
				\\&\,\,\,\,\,\,\,\,\,\,\,\,\,\,\,\,\,\,\,\,\,\,\,\,\,\,\,\,\,\,\,\,\,\,\,\,+{ \sum_{s=1}^{S}} Z^{s}e^{j2\pi (-\frac{2f_{c}+\Delta f(M-1)}{c})R^{s}}.
			\end{aligned}
		\end{split}
	\end{equation}
	
	The second term represents the static interference component, which remains unchanged across frames. As shown in (33), the variation of $R_{p}^{\delta}$ causes a change in the phase of $\mathbf{O}[p]$ per frame. Let us rewrite $\mathbf{O}[p]$ as 
	\begin{equation}
		\begin{split}\color{black}
			\begin{aligned}
				\mathbf{O}[p]=O_{\text{static}}+U_{\text{micro}}e^{j2\pi f_{\text{micro}}R_{p}^{\delta}} ,
			\end{aligned}
		\end{split}
	\end{equation}
	where $O_{\text{static}}= { \sum_{s=1}^{S}} Z^{s}e^{j2\pi (-\frac{2f_{c}+\Delta f(M-1)}{c})R^{s}}, U_{\text{micro}}=Z^{\delta}$ and $f_{\text{micro}}=-\frac{2f_{c}+\Delta f(M-1)}{c}$. 
	When static clutter has not yet been removed, the amplitude and phase of \( \mathbf{O}[p] \) in each frame describe a circle whose center is offset from the origin, as shown in Fig.~7(a). To obtain the circle center $O_{\text{static}}$, we perform the least squares circle fitting on $\mathbf{O}[:]$. 
	
	The effectiveness of circle fitting requires the assurance of the following conditions. Firstly, micro-deformation must be small (sub-centimeter scale, as shown in Fig.
	10), such that the rotating phasor's amplitude remains constant, approximating a circle.
	Secondly, SNR must exceed $-20$ dB to prevent noise from distorting the circular
	distribution. Additionally, CF-MSIR effectiveness requires channel coherence, which guarantees
	reliable phase characteristic identification and forms the foundation for deformation       
	measurement. Furthermore, CF-MSIR assumes weak multipath reflection effects. While
	multipath reflections can be analogous to multi-dynamic target interference, their
	impact is not considered in this work since we assume that deformation within the monitoring area
	exhibits consistency after beamforming.
	
	 Then, the micro-deformation effective echo component can be represented as 
	\begin{equation}
		\begin{split}\color{black}
			\begin{aligned}
				U_{\text{micro}}e^{j2\pi f_{\text{micro}}R_{p}^{\delta}}= \mathbf{O}[p]-O_{\text{static }}.
			\end{aligned}
		\end{split}
	\end{equation}
	
	As shown in Fig. 7(b), the CF-MSIR method effectively removes the static interference clutter in $\mathbf{O}[:]$, and obtains the micro-deformation effective echo component $\mathbf{O}_{\text{micro}}[p] = U_{\text{micro}}e^{j2\pi f_{\text{micro}}R_{p}^{\delta}}$. Then, the vector $\mathbf{O}_{\text{micro}}[p]$ can serve as the input for the subsequent section to estimate the bridge micro-deformation.

	\section{Micro-Deformation Monitoring Scheme}
	
	In this section, we propose a BMDM scheme, and present a phase unwrapping method based on historical micro-deformation data to ensure high-precision BMDM.	
	\subsection{The Proposed BMDM Scheme}
	Note that $\mathbf{O}_{\text{micro}}[p]$ can be expressed as
	\begin{equation}
		\begin{split}\color{black}
			\begin{aligned}
				\mathbf{O}_{\text{micro}}[p] = Z^{\delta}e^{j2\pi f_{\text{micro}}R_{p}^{\delta}}.
			\end{aligned}
		\end{split}
	\end{equation}

	Extracting the phase of $\mathbf{O}_{\text{micro}}[p]$, we obtain
	\begin{equation}
		\begin{split}\color{black}
			\begin{aligned}
				\varphi_{p}=2\pi f_{\text{micro} }R_{p}^{\delta} +\varphi_{Z^{\delta}}\,\,\,\,mod(2\pi ),
			\end{aligned}
		\end{split}
	\end{equation}
	where $\varphi_{p} \in (-\pi, \pi]$ is the phase of $\mathbf{O}_{\text{micro}}[p]$ in the $p$-th frame, and $\varphi_{Z^{\delta}}$ is the phase of $Z^{\delta}$. In fact, since \( R_p^d \) is generally large, the right-hand side of (37) may exceed the range \( (-\pi, \pi] \), which causes \( \varphi_p \) to be affected by phase wrapping. Unfortunately, existing phase unwrapping methods introduce significant errors in the measurement of $R_{p}^{\delta}$, which renders the direct determination of  $R_{p}^{\delta}$ from $\varphi_{p}$ infeasible. Nevertheless, we can utilize the phase difference $\Delta \varphi$ between $\mathbf{O}_{\text{micro}}[0]$ and $\mathbf{O}_{\text{micro}}[p]$ to obtain the micro-deformation, and $\Delta \varphi$ can be formulated as
	\begin{equation}
		\begin{split}\color{black}
			\begin{aligned}
				\Delta \varphi= \varphi_{p}-\varphi_{0}=2\pi  f_{micro}\Delta R^{\delta}_{0\to p},
			\end{aligned}
		\end{split}
	\end{equation}
	where \(\Delta R^{\delta}_{0\to p} = R_{p}^{\delta} - R_{0}^{\delta}\) denotes the radial change of the bridge micro-BMDM point at the $p$-th frame relative to the initial moment. Based on (38),  \(\Delta R^{\delta}_{0\to p}\) can be expressed as 
	\begin{equation}
		\begin{split}\color{black}
			\begin{aligned}
				\Delta R^{\delta}_{0\to p} = \frac{1}{2\pi f_{micro}}\Delta \varphi=\frac{-c}{2\pi (2f_{c}+\Delta f(M-1))}\Delta \varphi.
			\end{aligned}
		\end{split}
	\end{equation}
	
	\begin{figure}[!t]
		\centering
		\includegraphics[width=85mm]{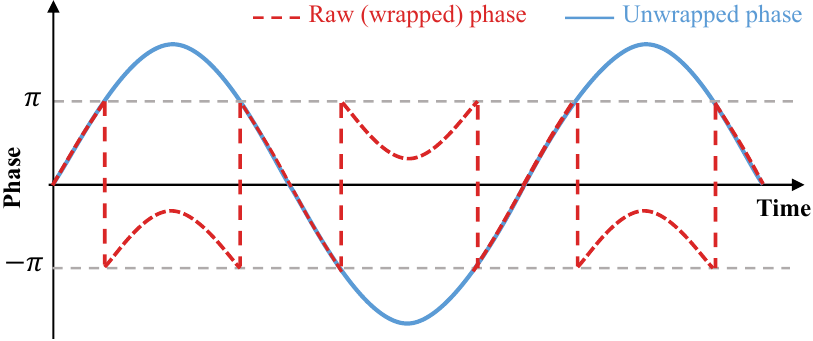}
		\caption{Simple example of the wrapped and unwrapped phases.}
		\label{fig_5}
	\end{figure}
	
	\subsection{The Proposed Phase Unwrapping Method}
	It should be noted that, as $R_{p}^{\delta}$ varies across different frames, $\varphi_{p}$ may experience a new phase wrapping at any time, i.e., a sudden jump from $\pi$ to $-\pi$ or from $-\pi$ to $\pi$, as shown in Fig.~8, which causes the measured \( \Delta R^d_{0\to p} \) to have an error of \( \frac{1}{f_{\text{micro}}} \) m.
	
	To address this issue, we propose a phase unwrapping method, which utilizes historical deformation data of the bridge to predict the unwrapped phase value. 
	\begin{figure*}[!t]
		\centering
		\includegraphics[width=180mm]{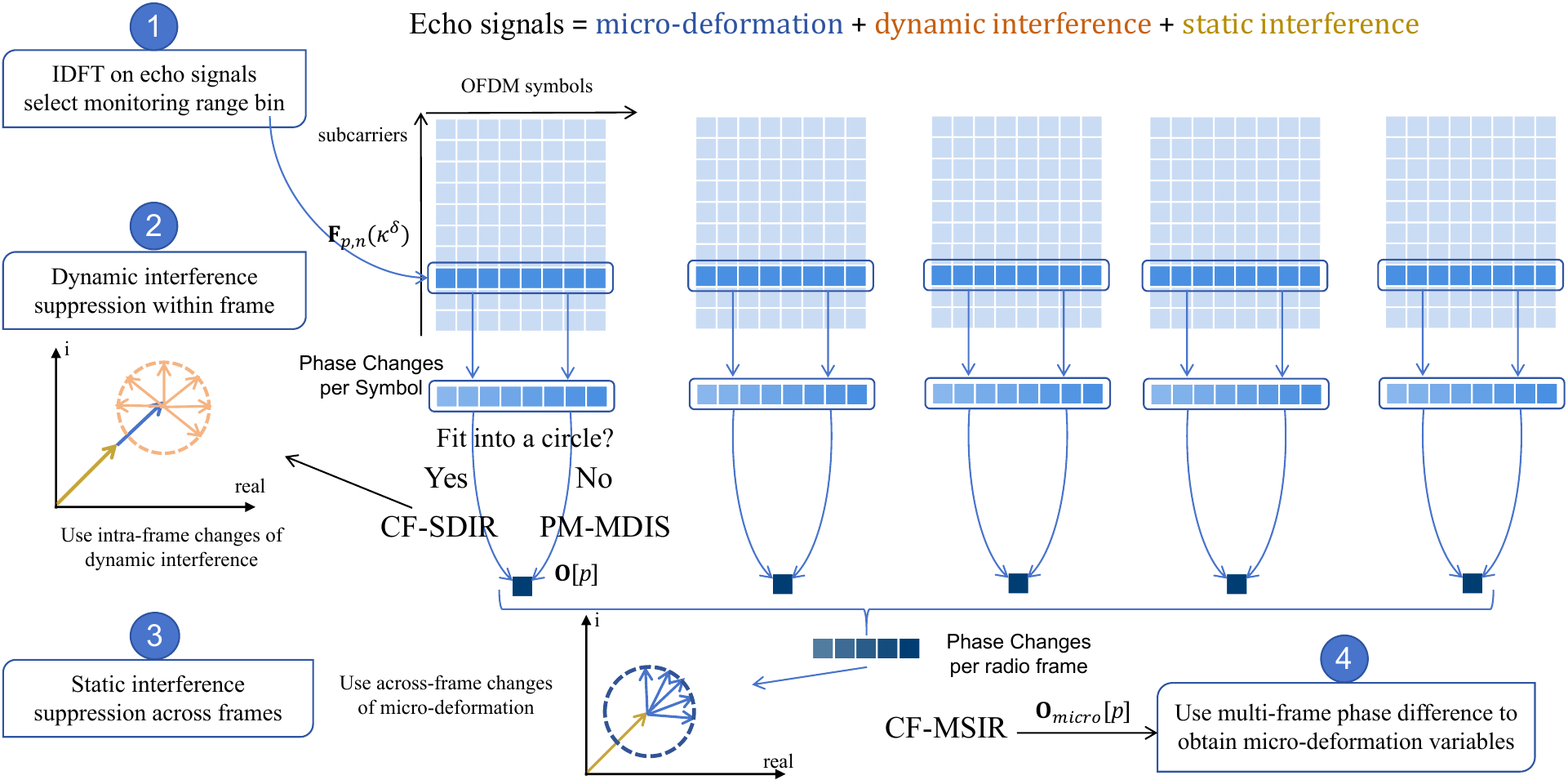}
		\caption{Flow chart of BMDM scheme.}
		\label{fig_4}
	\end{figure*}

	For clarity, assume that $\Delta R_{p}^{\delta}=\Delta R_{0\to p}^{\delta}$. Moreover, suppose no phase wrapping occurs at the first three sampling points, we could obtain the predicted distance as\cite{ma2023structural}
	\begin{equation}
		\begin{split}\color{black}
			\begin{aligned} 
				\Delta \hat{R}_{p}^{\delta}=2 \Delta R_{ p-1}^{\delta}-\Delta R_{ p-2}^{\delta}+(T_{f})^{2} a_{p-2},
			\end{aligned}
		\end{split}
	\end{equation}
	where $a_{p-2} = [\frac{\Delta R_{p -1}^{\delta}-\Delta R_{p -2}^{\delta}}{T_{f}}-\frac{\Delta R_{p -2}^{\delta}-\Delta R_{p -3}^{\delta}}{T_{f}}]/T_{f}$ represents the acceleration of the micro-BMDM point in the $(p\!-\!2)$-th frame. Then, the phase variation in the $p$-th frame can be predicted as
	\begin{equation}
		\begin{split}\color{black}
			\begin{aligned} 
				\hat{\varphi}_{p }=2\pi f_{\text{micro}}\Delta \hat{R}_{p}^{\delta} .
			\end{aligned}
		\end{split}
	\end{equation}
	
	According to the modulo $2\pi$ principle in formula (37), the value of the unwrapped phase could be one of the $\bar{\varphi}_{p}$ whose fomula is
	\begin{equation}
		\begin{split}\color{black}
			\begin{aligned} 
				\bar{\varphi}_{p}=\varphi_{p } \pm 2 \gamma     \pi,
			\end{aligned}
		\end{split}
	\end{equation}
	where $\gamma    \in \mathbb{Z}$ is an integer. Moreover,
	the unwrapping phase $\bar{\varphi}_{p}$ is estimated as the one that is the closest to the predicted phase $\hat{\varphi}_{p }$, and can be represented  as
	\begin{equation}
		\begin{split}\color{black}
			\begin{aligned} 
				\bar{\varphi}_{p }=\varphi_{p}+2 \pi \times 
				\left \lfloor  \frac{\hat{\varphi}_{p}-\varphi_{p}}{2 \pi} \right \rceil.
			\end{aligned}
		\end{split}
	\end{equation}
	
	Then, the radial distance change of the bridge micro-BMDM point in the $p$-th frame after phase unwrapping is expressed as
	\begin{equation}
		\begin{split}\color{black}
			\begin{aligned} 
				\Delta R_{p }^{\delta}=\frac{1}{2\pi f_{micro}}\left(\bar{\varphi}_{p }-\varphi_{0}\right).
			\end{aligned}
		\end{split}
	\end{equation}
	
	The vertical micro-deformation $\Delta D_{0\to p}$ of the bridge monitoring point can be successfully estimated by the coordinate transformation whose formula is
	\begin{equation}
		\begin{split}\color{black}
			\begin{aligned} 
				\Delta D_{0\to p} = z^{\delta}_{0}-\sqrt{(z^{\delta}_{0})^{2}+(\Delta R_{p }^{\delta})^{2}-2R_{0}^{\delta}\Delta R_{p}^{\delta}} .
			\end{aligned}
		\end{split}
	\end{equation}
	
	The general flow chart of interference suppression and BMDM scheme are shown in Fig.~9. First, we perform an IDFT on the echo signal in the subcarrier dimension to preliminarily extract the micro-deformation feature vector. Next, since the number of dynamic interference on the bridge is unknown, we directly apply circle fitting to the feature vector. If the data points conform to a circular distribution, then the CF-SDIR method is employed to remove dynamic clutter. Otherwise, the PM-MDIS method is utilized to suppress dynamic clutter. Subsequently, we develop the CF-MSIR method to remove static clutter. The micro-deformation variable of the bridge is calculated based on the multi-frame phase difference after phase unwrapping.
	\subsection{Computational Complexity Analysis}
	The computational complexity of the proposed BMDM scheme consists of three components: IDFT operations, clutter suppression, and micro-deformation estimation. Specifically, the computational complexity of IDFT operations is $O\{G(PNM^{2})\}$, the 
	complexity of clutter suppression is $O\{G(PN+P)\}$, and the complexity of micro-deformation estimation is $O\{G(P)\}$. This analysis demonstrates that the computational complexity of the proposed BMDM scheme is dominated by the IDFT operations. The overall 
	complexity satisfies real-time processing requirements and enables straightforward deployment on edge devices.

	\section{Simulation Results}
	
	In simulations, we set the number of antennas in HU-UPA as $N_{H}^{x} \times N_{H}^{z} = $ $8$ $\times$ $8$, the number of antennas in RU-UPA as $N_{R}^{x} \times N_{R}^{z} =$ $8$ $\times$ $8$, the carrier frequency as $f_{c} =$ $26$~GHz, the antenna spacing as $d_{x}=d_{z}=d=\frac{\lambda}{2}=\frac{c}{2{\small f_{c}}}$, and the subcarrier spacing as $\Delta f =$ $480$~kHz. For the 5G NR frame structure, we set the total number of frames as $P = 1500$, the duration of a frame as $T_{f} =$ $10$~ms, the corresponding total monitoring duration as $15$ s, the duration of one PRI (``MMMSU'' or ``DDDSU'') as $T_{c} =$ $2.5$~ms, the duration of a functional time slot as $0.5$~ms.  Moreover, the duration of one OFDM symbol is set as $T_{\text {sym}} =$ $10$~us. The adaptive threshold for the CPM method is set to $0.2$ times the fitted radius, and the	distribution is determined to be circular when the proportion of qualified points
	exceeds $90\%$. The noise is assumed to obey the complex Gaussian distribution with mean $\mu=0$ and  variance $\sigma_n^2=1$. 
	
	To evaluate the performance of the proposed BMDM scheme, we define the root mean square error (RMSE) of bridge micro-deformation as
	\begin{equation}
		\begin{split}\color{black}
			\begin{aligned}
				\mathrm{RMSE}_{D }=\sqrt{\frac{\sum_{\varsigma  =1}^{\text {Count }}\sum_{p=1}^{P} \left(\Delta\hat{D }_{\varsigma ,p}-\Delta D ^{truth}_{\varsigma ,p}\right)^{2}}{\!\!\!\text{Count}\times P}},
			\end{aligned}
		\end{split}
	\end{equation}
	where $Count$ is the number of the Monte Carlo runs, $\Delta D ^{truth}_{\varsigma ,p}$ and $\Delta \hat{D }_{\varsigma ,p}$ are the true value and the estimated value of the micro-deformation in the $p$-th frame of the $\varsigma$-th repeated trial, respectively.  Besides, the signal-to-noise power ratio (SNR) is defined as
	\begin{equation}
		\begin{split}\color{black}
			\begin{aligned}
				\mathrm{SNR}=\frac{\mathbb{E}\left\{\left|\mathbf{w}^{H} \mathbf{H}_{p,n, m}^{\text {channel }} \mathbf{x}_{p,n, m}^{*}\right|^{2}\right\}}{\mathbb{E}\left\{\left|n_{p,n, m}\right|^{2}\right\}},
			\end{aligned}
		\end{split}
	\end{equation}
	where $\mathbb{E}\left\lbrace \cdot  \right\rbrace $ represents the mathematical expectation operator.
	It is worth pointing out that the existing ISAC schemes that do not consider  dynamic and static clutter are unable to perform BMDM in actual clutter environment. Thus, we did not show the performance of the existing methods in the simulations.
	
	\subsection{The Generation of Bridge Micro-Deformation Data}
	\begin{table}[H]
		\centering
		\caption{Bridge Modeling Parameters}
		\label{tab:bridge_params}
		\setlength{\tabcolsep}{13pt} % 调整列间距
		\renewcommand{\arraystretch}{1.7} % 调整行高
		\begin{tabular}{|c|c|c|}
			\hline
			\textbf{Symbol} & \textbf{Parameter} & \textbf{Value} \\ \hline
			\(W\) & Length of the bridge &  100 m \\ \hline
			\(\xi\) & Young's modulus &  \(2.943 \times 10^{10}\) Pa \\ \hline
			\(I_B\) & Moment of inertia &  8.65 \(\text{m}^4\) \\ \hline
			\(\rho _{B}\) & Mass per unit length &  \(3.6 \times 10^4\) kg/m \\ \hline
			\(\zeta\) & Damping factor &  -0.02 \\ \hline
		\end{tabular}
	\end{table}
	\begin{table}[H]
		\centering
		\caption{Excitation Source Parameters}
		\label{tab:excitation_params}
		\setlength{\tabcolsep}{12pt} % 调整列间距
		\renewcommand{\arraystretch}{1.7} % 调整行高
		\begin{tabular}{|c|c|c|c|}
			\hline
			\textbf{\(Source_e\)} & \textbf{ (\(A_e\))} & \textbf{ (\(f_e\))} & \textbf{  (\( L_{e}\))} \\ \hline
			1 & 5.15 mm & 0.8 Hz & (145,60,-25) m \\ \hline
			2 & 3.72 mm & 1.5 Hz & (165,60,-25) m \\ \hline
			3 & 4.59 mm & 0.6 Hz & (180,60,-25) m \\ \hline
			4 & 3.42 mm & 1.1 Hz & (195,60,-25) m \\ \hline
			5 & 4.68 mm & 1.3 Hz & (220,60,-25) m \\ \hline
		\end{tabular}
	\end{table}
	\begin{figure*}[!t]
		\centering
		\subfloat[]{\includegraphics[width=60mm]{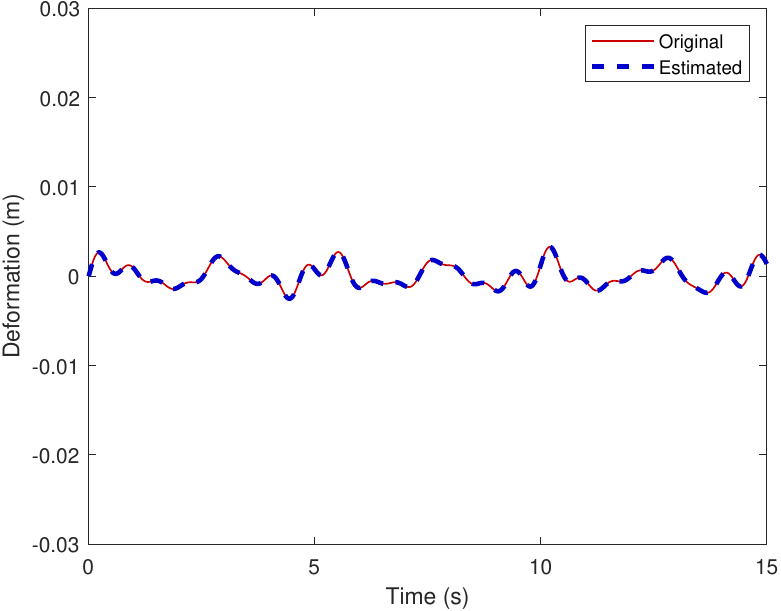}%
			\label{fig_first_case}}
		\hspace{-2mm} % 调整这里的数值来增加或减少间距
		\hfil
		\subfloat[]{\includegraphics[width=60mm]{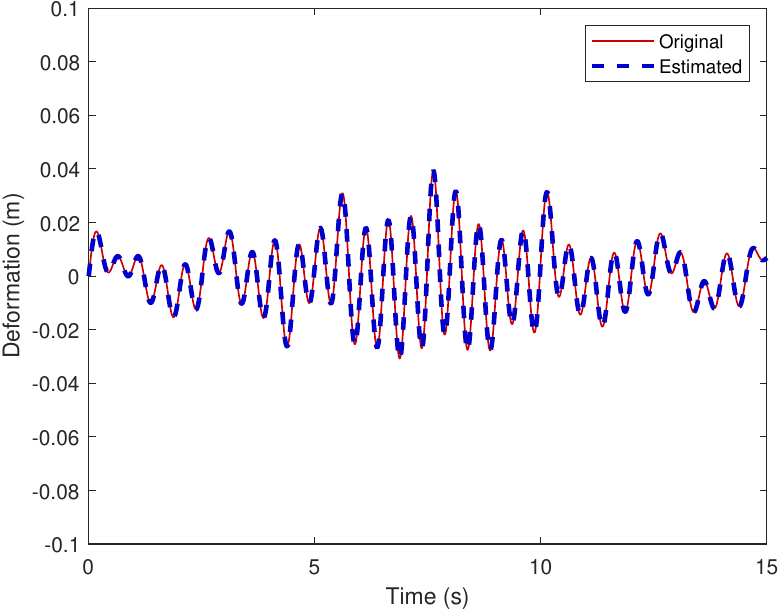}%
			\label{fig_second_case}}
		\hspace{-2mm} % 调整这里的数值来增加或减少间距
		\hfil
		\subfloat[]{\includegraphics[width=60mm]{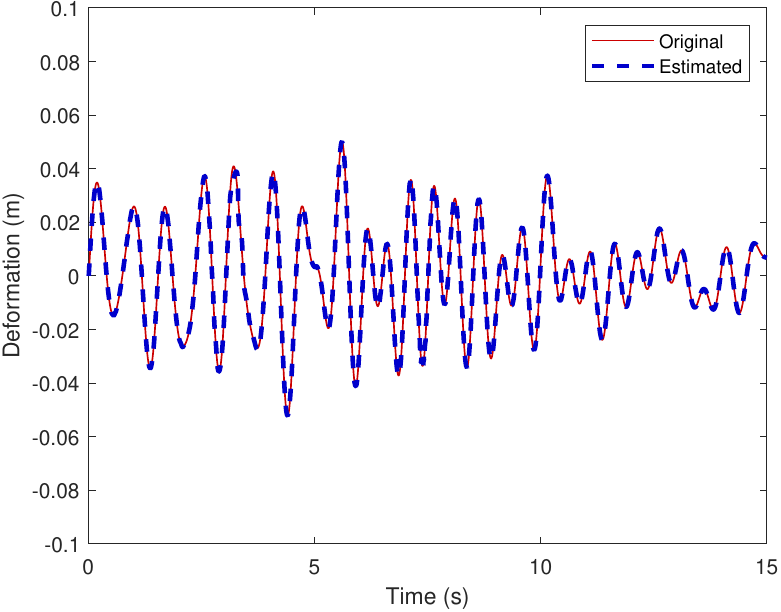}%
			\label{fig_second_case}}
		\caption{(a) The BMDM results in Condition \RNum{1}. 
			(b) The BMDM results in Condition \RNum{2}. (c) The BMDM results in Condition \RNum{3}.  \textcolor{black}{The BMDM results under different conditions.}}
		\label{fig_sim}
	\end{figure*}
	\begin{figure*}[!t]
		\centering
		\subfloat[]{\includegraphics[width=60mm]{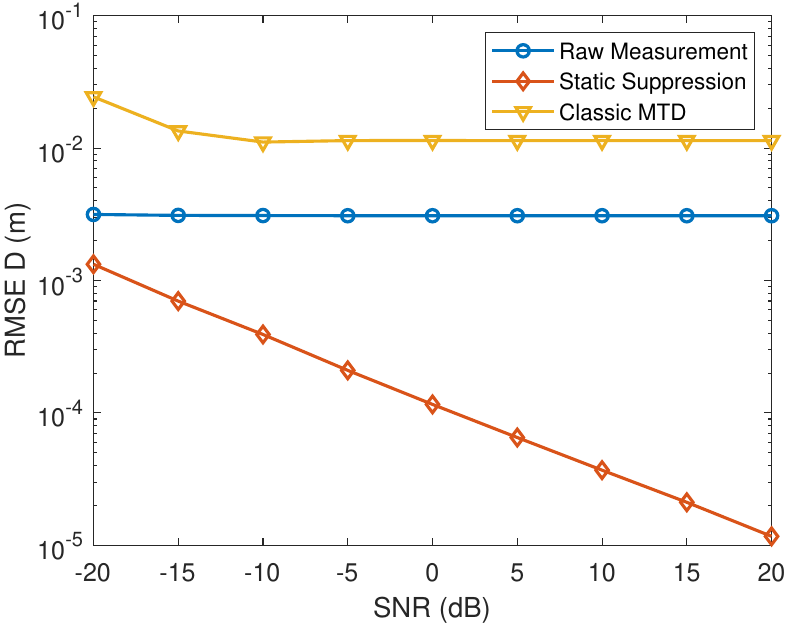}%
			\label{fig_first_case}}
		\hspace{-2mm} % 调整这里的数值来增加或减少间距
		\hfil
		\subfloat[]{\includegraphics[width=60mm]{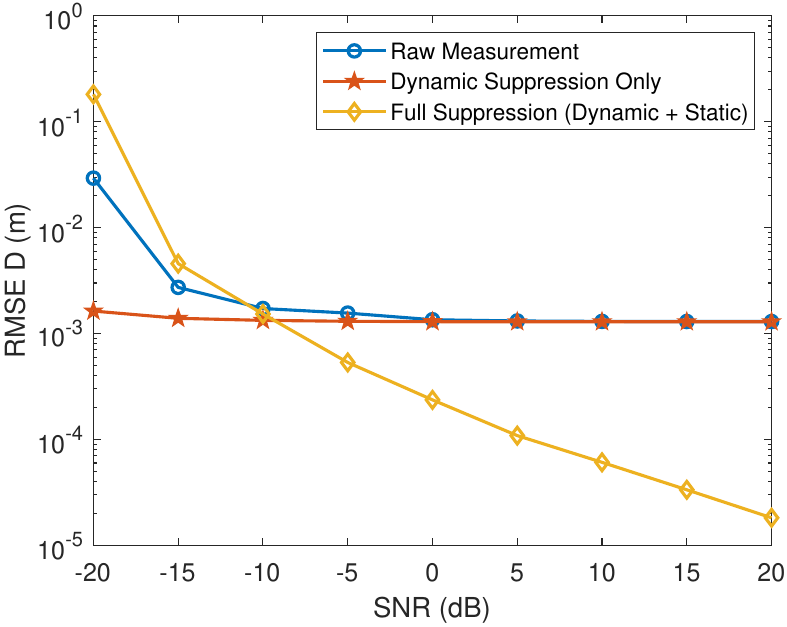}%
			\label{fig_second_case}}
		\hspace{-2mm} % 调整这里的数值来增加或减少间距
		\hfil
		\subfloat[]{\includegraphics[width=60mm]{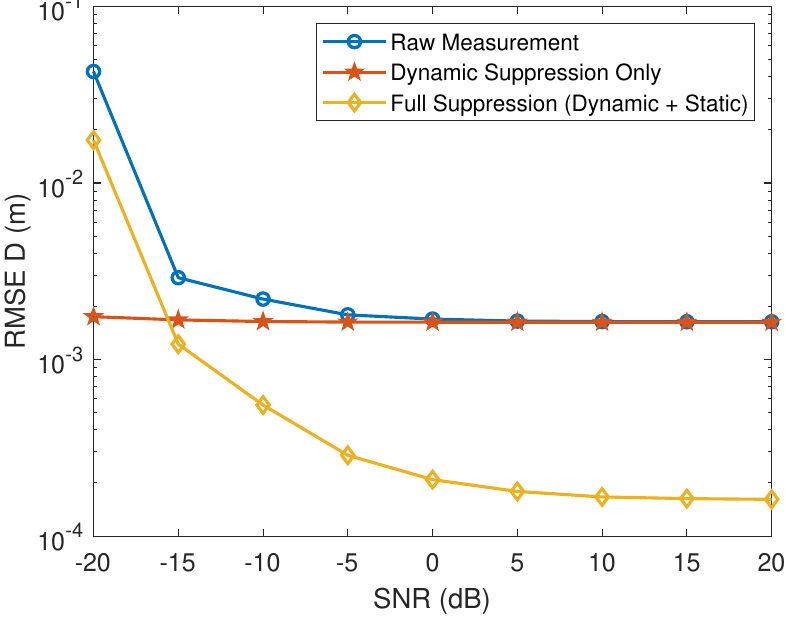}%
			\label{fig_second_case}}
		\caption{(a) Impact of static interference suppression on monitoring performance in Condition \RNum{1}. 
			(b) Impact of dynamic interference suppression and static interference suppression on monitoring performance in Condition \RNum{2}. (c) Impact of dynamic interference suppression and static interference suppression on monitoring performance in Condition \RNum{3}.  }
		\label{fig_sim}
	\end{figure*}
	
	In bridge modeling, we set the amplitude and initial phase  of the free deformation as $A_{0} = 1.35$ mm and  $\varphi_{B} = 0$, respectively. We set the direction vector of the bridge as $\vec{l}=(1,0,0)$, and the BMDM point is set at the mid-span of the bridge with the coordinate $  (x^{\delta}_{0}, y^{\delta}_{0}, z^{\delta}_{0})=(180,60,-25)$~m. The inherent parameters of the bridge are specified in Table~\RNum{2}. Then, we can calculate the bridge fundamental frequency as $f_{B}=\frac{\pi}{2 W^{2}} \sqrt{\frac{\xi I_{B}}{\rho _{B}}} = 0.42$ Hz. We fixed the parameters of $5$ static excitation sources, as shown in Table~\RNum{3}. 
	The dynamic excitation sources are generated in the form of dynamic interferences, which not only affects the deformation of the bridge but also distorts the echo signals. We randomly set $S$ static interferences in the BS workspace and $K$ dynamic interferences on the bridge. Due to the different masses and velocities of the dynamic interferences, their effect on the bridge is difficult to predict. The amplitude and frequency of the dynamic excitation sources are randomly set from $10$ mm to $50$ mm and $0.2$ Hz to $5$ Hz. The initial phases of static and dynamic excitation sources are set to $0$. For clarity, we set three simulation conditions that are defined as:
	$E = 5$, $S = 8$, and $K = 0$ (Condition \RNum{1});
	$E = 5$, $S = 8$, and $K = 1$ with a speed of $12$ m/s along the bridge (Condition \RNum{2});
	$E = 5$, $S = 8$, and $K = 3$ with speeds along the bridge of $30$ m/s, $15$ m/s, and $-12$ m/s, respectively (Condition \RNum{3}).
	
	Based on (1), (2), and (3), we could obtain  the micro-deformation of the bridge, as shown in Fig.~10. Fig.~10(a), Fig.~10(b), and Fig.~10(c) show the deformation of the bridge monitoring point in Condition \RNum{1}, Condition \RNum{2}, and Condition \RNum{3}, respectively. We observed that bridge micro-deformations generally occur at the centimeter scale, which necessitates sub-millimeter measurement accuracy to reliably perform BMDM.
	Then, we could visualize the performance of the bridge micro-deformation scheme by demonstrating the RMSE-SNR curve.

	\begin{figure*}[!t]
		\centering
		\subfloat[]{\includegraphics[width=60mm]{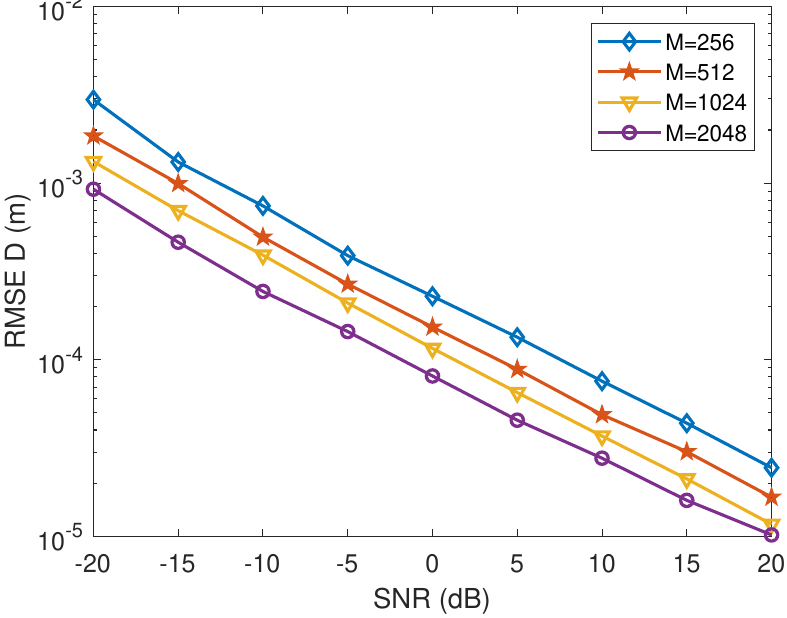}%
			\label{fig_first_case}}
		\hspace{-2mm} % 调整这里的数值来增加或减少间距
		\hfil
		\subfloat[]{\includegraphics[width=60mm]{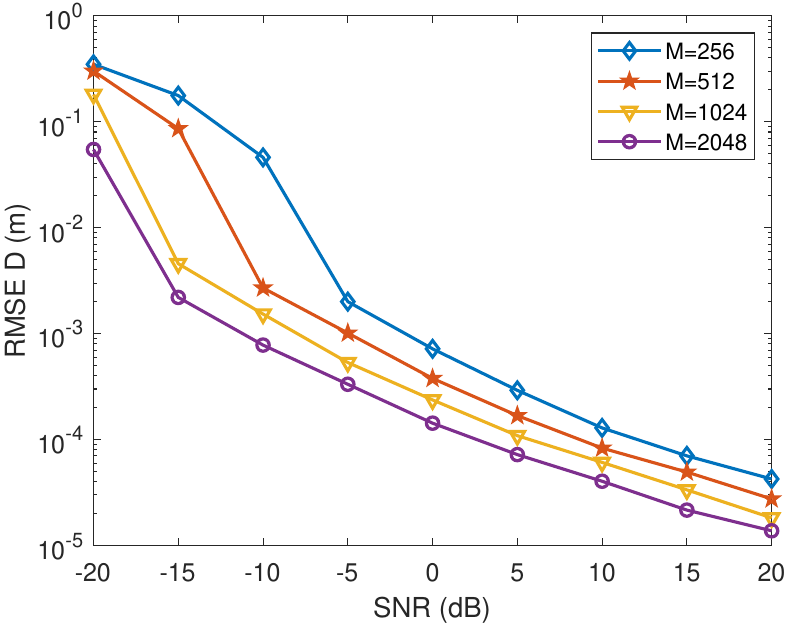}%
			\label{fig_second_case}}
		\hspace{-2mm} % 调整这里的数值来增加或减少间距
		\hfil
		\subfloat[]{\includegraphics[width=60mm]{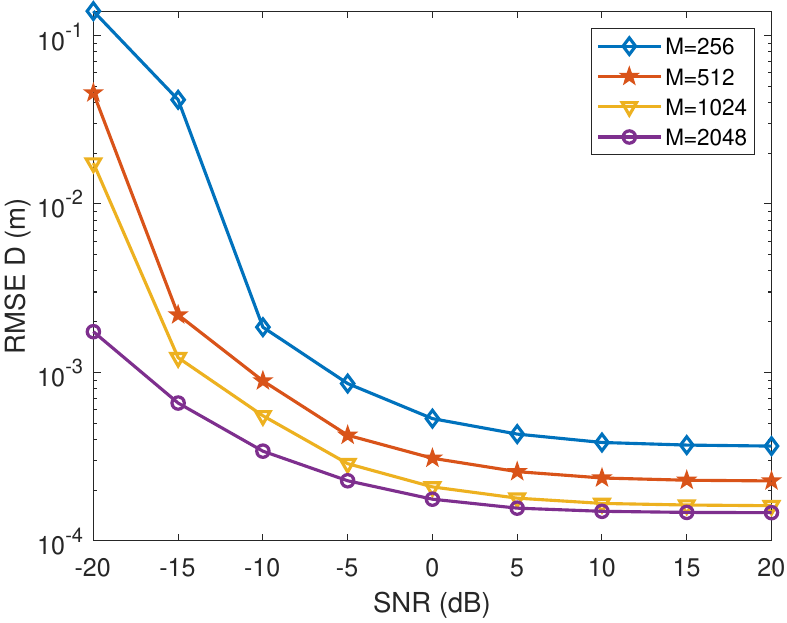}%
			\label{fig_second_case}}
		\caption{(a) BMDM performance under different numbers of subcarriers in Condition \RNum{1}. 
			(b) BMDM performance under different numbers of subcarriers in Condition \RNum{2}. (c) BMDM performance under different numbers of subcarriers in Condition \RNum{3}.  }
		\label{fig_sim}
	\end{figure*}
	\begin{figure*}[!t]
		\centering
		\subfloat[]{\includegraphics[width=60mm]{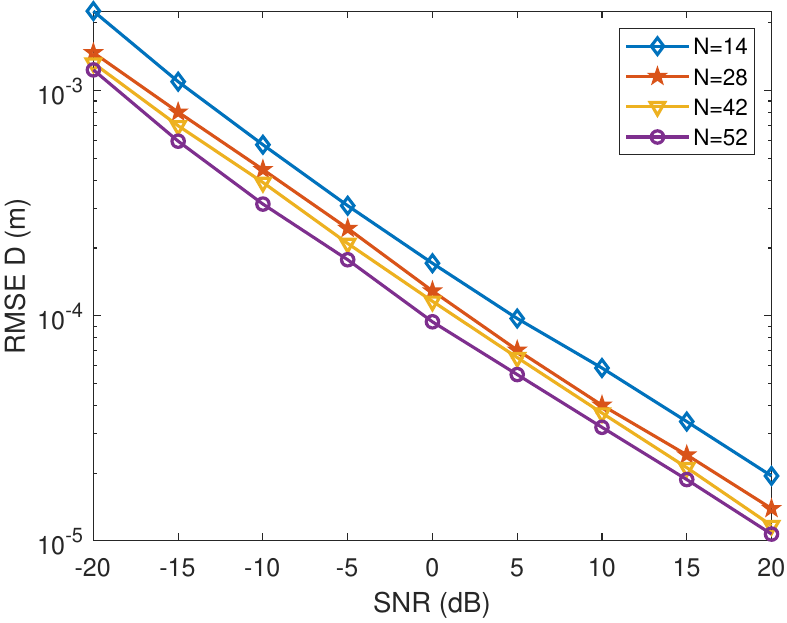}%
			\label{fig_first_case}}
		\hspace{-2mm} % 调整这里的数值来增加或减少间距
		\hfil
		\subfloat[]{\includegraphics[width=60mm]{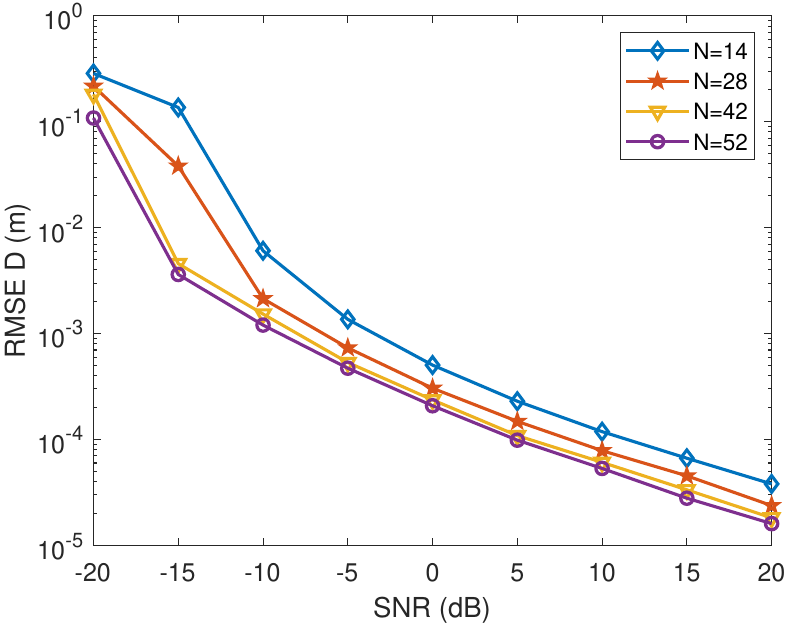}%
			\label{fig_second_case}}
		\hspace{-2mm} % 调整这里的数值来增加或减少间距
		\hfil
		\subfloat[]{\includegraphics[width=60mm]{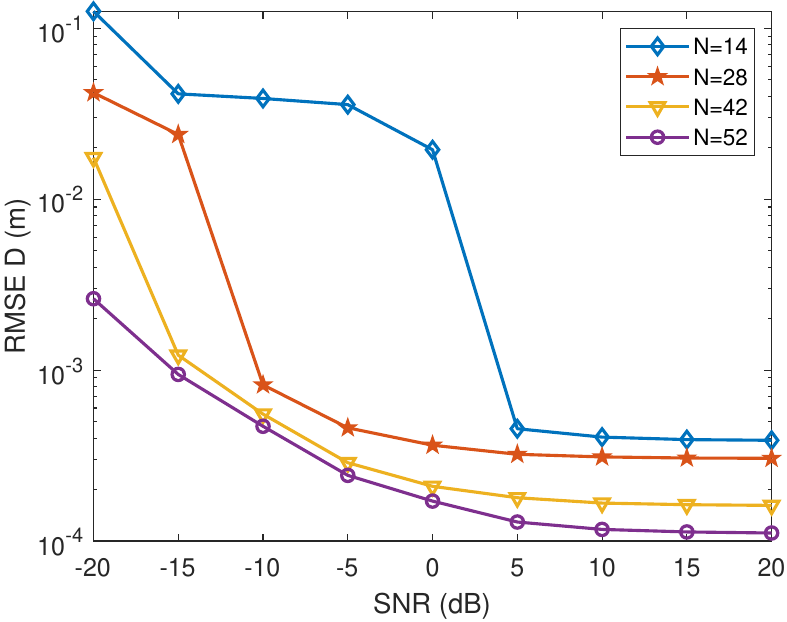}%
			\label{fig_second_case}}
		\caption{(a) BMDM performance under different numbers of OFDM symbols in Condition \RNum{1}. 
			(b)  BMDM performance under different numbers of OFDM symbols in Condition \RNum{2}. (c) BMDM performance under different numbers of OFDM symbols in Condition \RNum{3}.  }
		\label{fig_sim}
	\end{figure*}
	\subsection{The Performance of BMDM Scheme}
	We set the number of subcarriers  $M = 1024$ and utilize $N = 42$ OFDM symbols for BMDM. The BMDM performance in dynamic and static clutter environment is affected by the interference conditions, and the results are shown in Fig.~11.

	Fig.~11(a) shows the variation of RMSE versus SNR for BMDM in Condition \RNum{1}.
	As the SNR increases, the monitoring accuracy is unable to improve without removing the static interference. The classic Moving Target Detection (MTD) techniques perform poorly in this situation, as they suppress static clutter while simultaneously attenuating the energy of micro-deformation targets. After utilizing the CF-MSIR method, $\mathrm{RMSE}_{D}$ continues to decrease, which indicates that the static interferences have been completely eliminated. When the SNR increases to $-15$ dB, the $\mathrm{RMSE}_{D}$ decreases to $10^{-3}$~m, which illustrates that the BMDM scheme can achieve sub-millimeter accuracy.
	
	Fig.~11(b) shows the variation of RMSE versus SNR for BMDM in Condition \RNum{2}. It can be observed that without the removal of dynamic interferences and static interferences, it is unable to achieve sub-millimeter accuracy. We could employ the CF-SDIR method to remove the dynamic clutter and perform the CF-MSIR method to remove the static clutter. When the SNR increases to $-5$ dB, the $\mathrm{RMSE}_{D}$ decreases to $10^{-3}$ m, which proves that the BMDM scheme could achieve sub-millimeter accuracy in this case. As the SNR continues to increase, the $\mathrm{RMSE}_{D}$ further decreases, which indicates that both dynamic interference and static interferences have been completely eliminated.
	
	Fig.~11(c) shows the variation of RMSE versus SNR for BMDM in Condition \RNum{3}. We first utilize the PM-MDIS method to suppress the dynamic clutter. Then, we employ the CF-MSIR method to remove the static clutter. With the suppression of the clutter in the environment, the proposed BMDM scheme achieves sub-millimeter accuracy as the SNR increases to $-10$ dB. Due to the presence of the dynamic interference margin $\epsilon_{p}^{dynamic}$, the trend of $\mathrm{RMSE}_{D}$ reduction slows down and eventually flattens as the SNR continues to increase. When the SNR increases to 20 dB, the RMSE decreases to $\mathrm{RMSE}_{D} = 0.00012$ m, which illustrates that while PM-MDIS is unable to completely remove dynamic interferences, it still achieves satisfactory suppression.

	\subsection{The Impact of Subcarriers on BMDM Scheme}
We utilize $N = 42$ OFDM symbols for BMDM and explore the effect of BMDM with different numbers of subcarriers, as shown in Fig.~12.

	Fig.~12(a) shows the variation of RMSE versus SNR for BMDM under different numbers of subcarriers in Condition \RNum{1}. It can be observed that the $\mathrm{RMSE}_{D}$ significantly decreases with the increase of the number of subcarriers $M$. In fact, a larger $M$ means a larger bandwidth, which increases the number of points for circle fitting in the CF-MSIR method to improve the performance of the BMDM method.
	
	Fig.~12(b) shows the variation of RMSE versus SNR for BMDM under different numbers of subcarriers in Condition \RNum{2}. The $\mathrm{RMSE}_{D}$ gradually decreases as the number of subcarriers $M$ increases. Based on (39), we could infer that the increase of $M$ improves the resolution of BMDM. By increasing $M$ to 2048, the BMDM scheme could achieve sub-millimeter accuracy when SNR increases to $-10$ dB.
	
	Fig.~12(c) shows the variation of RMSE versus SNR for BMDM under different numbers of subcarriers in Condition \RNum{3}. When $M$ increases to $2048$, the RMSE decreases to $\mathrm{RMSE}_{D} = 0.00015$ m at SNR $= 20$ dB, which indicates that the proposed BMDM scheme has high monitoring accuracy in complex clutter environment. This is because a larger $M$ enhances the clutter isolation effect of the IDFT method, which makes the micro-deformation echo less susceptible to dynamic interferences and static interferences.

	\subsection{The Impact of OFDM Symbols on BMDM Scheme}

	We set the number of subcarriers  $M = 1024$ for BMDM and explore the effect of BMDM with different numbers of OFDM symbols, as shown in Fig.~12. To comply with the 5G NR protocol, we set $N = 14$, $N = 28$, $N = 42$, and $N = 52$ for simulations.

	Fig.~13(a) shows the variation of RMSE versus SNR for BMDM under different numbers of OFDM symbols in Condition \RNum{1}. It can be seen from Fig.~13(a) that the $\mathrm{RMSE}_{D}$ significantly decreases with the increase of the number of OFDM symbols $N$. This is because more $N$ provides more phase resources, which enhances the performance of the BMDM scheme.
	
	Fig.~13(b) shows the variation of RMSE versus SNR for BMDM under different numbers of OFDM symbols in Condition \RNum{2}. As the $N$ increases, the $\mathrm{RMSE}_{D}$ gradually decreases. Based on (27), the increase of $N$ could improves the number of fitting points for CF-SDIR, which enhances the overall performance of the BMDM scheme.
	
	Fig.~13(c) shows the variation of RMSE versus SNR for BMDM under different numbers of OFDM symbols in Condition \RNum{3}. When $N$ increases to $52$, the RMSE decreases to $\mathrm{RMSE}_{D} = 0.00011$ m at SNR $= 20$ dB, which indicates that the proposed BMDM scheme has high monitoring accuracy in complex clutter environment. In fact, based on (31) and (32), the increase of $N$ not only improves the effectiveness of PM-MDIS but also suppresses noise, which leads to an increase in the accuracy of BMDM scheme.
	\begin{figure}[!t]
		\centering
		\includegraphics[width=80mm]{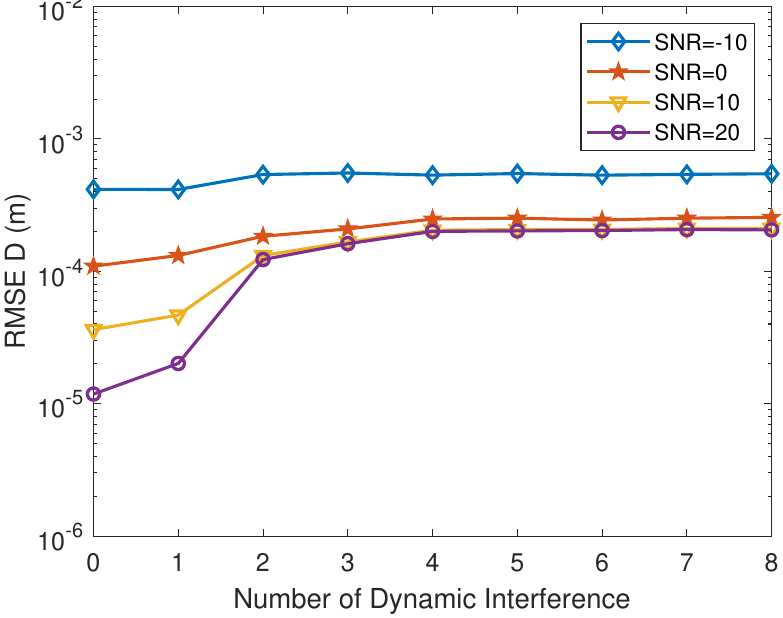}
		\caption{The curves of RMSE versus number of dynamic interference under different SNR.}
		\label{fig_5}
	\end{figure}
	\subsection{The Impact of Different Number of Dynamic Interference on BMDM Scheme}
	We set the number of subcarriers  $M = 1024$ and utilize $N = 42$ OFDM symbols for BMDM. Then, we randomly set $S = 8$ static interferences in the BS workspace.
	To further explore the impact of the number of dynamic interference on BMDM, we set $K$ dynamic interferences on the bridge with random starting positions and random speeds  that range from $0$ m/s to $50$ m/s along the bridge, where $K$ is set from $0$ to $8$. 
	
	Fig.~14 shows the variation of RMSE versus the number of dynamic interferences for BMDM under different SNRs. 
	When $K = 0$, static interference is the main source of interference, and the performance of the BMDM scheme is affected by the circle fitting of multi-frame data. As SNR increases from $-10$ dB to $10$ dB, $\mathrm{RMSE}_{D}$ decreases from $0.00042$ m to $0.000012$ m.
	When $K$ = 1, a single dynamic interference is introduced on the basis of multiple static clutters, and the two circle fittings of the CF-SDIR and CF-MSIR methods cause a slight increase in $\mathrm{RMSE}_{D}$.
	Based on (32), the PM-MDIS method can only suppress but not completely remove dynamic clutter in the presence of multiple dynamic interferences. As $K$ increases from $2$ to $8$, the RMSE continues to rise. Nevertheless, when the SNR is greater than $0$ dB, sub-millimeter accuracy is achieved regardless of the number of dynamic interferences, which demonstrates the effectiveness and robustness of the proposed BMDM scheme.
	
	\subsection{Evaluation of the Effectiveness of PM-MDIS}
	We set the number of subcarriers  $M = 1024$ for BMDM to explore the effectiveness of PM-MDIS, as shown in Fig.~15. 
	\begin{figure}[!t]
		\centering
		\includegraphics[width=80mm]{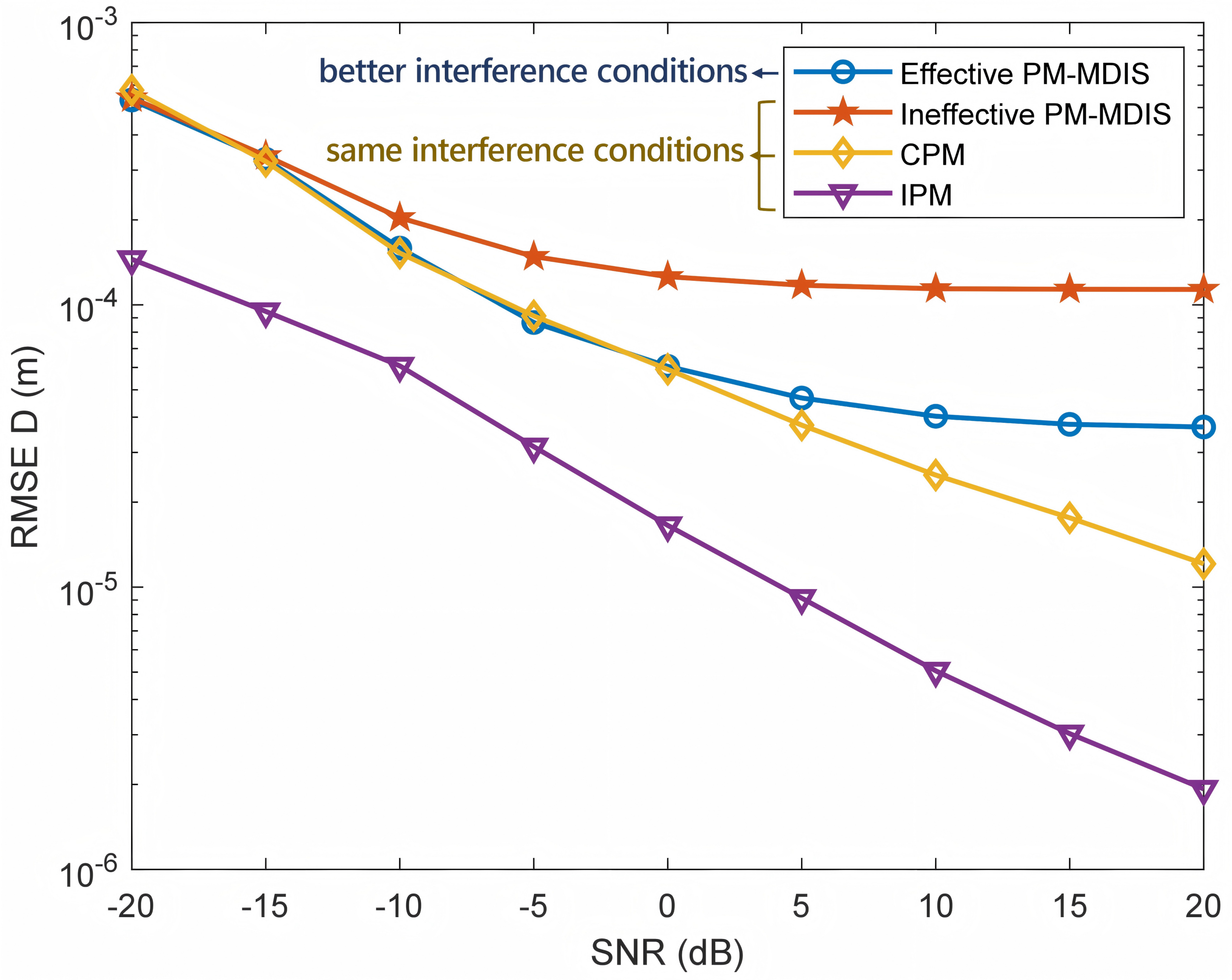}
		\caption{The performance of dynamic interference algorithms under different interference conditions.}
		\label{fig_5}
	\end{figure}
	
	From the PM-MDIS effectiveness condition formula $v_{p}^{k} = \gamma_{p}^{k} \cdot
	\frac{c}{2f_0NT_s}$, when $N = 28$, there is $v_{p}^{k} \approx 20$ m/s.
	Therefore, in the ``effective PM-MDIS'' experiment, we set a dynamic interference with a      
	radial velocity of $20$ m/s when passing near the monitoring point. In the ``ineffective       
	PM-MDIS'' experiment, we set a dynamic interference with a radial velocity of 10 m/s when    
	passing near the monitoring point. It can be observed that PM-MDIS achieves better
	suppression performance for interference with a velocity of 20 m/s. We can also observe     
	that the CPM method, which jointly employs CF-SDIR, shows significantly improved
	interference suppression performance. This is because CPM automatically switches to the     
	CF-SDIR method after detecting the single dynamic interference data distribution. All       
	the aforementioned methods use $N = 28$ OFDM symbols. The IPM method uses a number of       
	OFDM symbols much larger than $28$ to ensure that the condition $v_{p}^{k} =
	\gamma_{p}^{k} \cdot \frac{c}{2f_0NT_s}$ is always satisfied, thus achieving the best       
	performance.
	
\subsection{The impact of beam misalignment on BMDM scheme}
We set the number of subcarriers  $M = 1024$ and utilize $N = 42$ OFDM symbols for BMDM.
	
	\begin{figure}[!t]
		\centering
		\includegraphics[width=80mm]{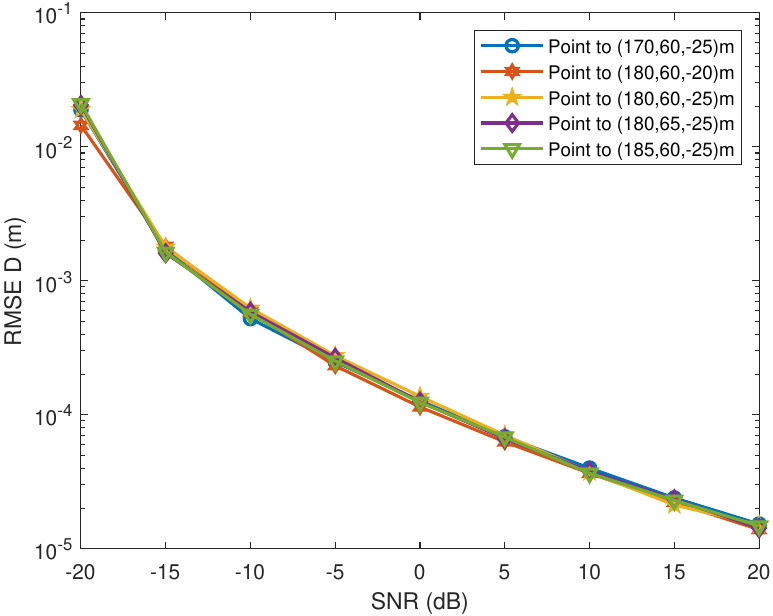}
		\caption{The performance of BMDM scheme under different beam misalignment conditions.}
		\label{fig_5}
	\end{figure}
	
	Fig.~16 shows the variation of RMSE versus SNR for BMDM under different beam misalignment conditions in Condition \RNum{3}. The simulation results demonstrate that the proposed BMDM scheme can still
		achieve sub-millimeter accuracy when the beam misalignment is not severe (offset
		distance $< 10$ m), with performance comparable to that of perfect alignment. This
		demonstrates the robustness of the BMDM scheme against beam misalignment effects.

\section{Conclusions}
	
	In this paper, we have proposed a novel BMDM scheme with ISAC system to monitor the bridge in complex interference conditions. We have provided an excitation-bridge coupling model to represent the micro-deformation process of the bridge and constructed the OFDM echo channel model for basic scene of BMDM that includes micro-deformation, dynamic objects and static environment.  To achieve high-precision monitoring, we have developed CF-SDIR and PM-MDIS to eliminate dynamic interferences in frame. Meanwhile, we have presented CF-MSIR to remove static interferences across multiple frames. Furthermore, we have leveraged the time-frequency phase resources of OFDM signals to perform BMDM and solved the problem of phase wrapping. Simulation results have demonstrated the effectiveness and robustness of the proposed scheme, which confirms a new finding that ISAC system can achieve cost-effective, high-precision, and highly stable BMDM.

	%\balance
	\bibliographystyle{ieeetr}
	\bibliography{this.bib}

\begin{thebibliography}{10}

\bibitem{zonzini2020structural}
F.~Zonzini, C.~Aguzzi, L.~Gigli, L.~Sciullo, N.~Testoni, L.~De~Marchi,
  M.~Di~Felice, T.~S. Cinotti, C.~Mennuti, and A.~Marzani, ``Structural health
  monitoring and prognostic of industrial plants and civil structures: A sensor
  to cloud architecture,'' {\em IEEE Instrum. Meas. Mag.}, vol.~23, no.~9,
  pp.~21--27, Dec. 2020.

\bibitem{di2021structural}
F.~Di~Nuzzo, D.~Brunelli, T.~Polonelli, and L.~Benini, ``Structural health
  monitoring system with narrowband {I}o{T} and {MEMS} sensors,'' {\em IEEE
  Sens. J.}, vol.~21, no.~14, pp.~16371--16380, Apr. 2021.

\bibitem{dutta2021recent}
C.~Dutta, J.~Kumar, T.~K. Das, and S.~P. Sagar, ``Recent advancements in the
  development of sensors for the structural health monitoring ({SHM}) at
  high-temperature environment: A review,'' {\em IEEE Sens. J.}, vol.~21,
  no.~14, pp.~15904--15916, Apr. 2021.

\bibitem{zonzini2020vibration}
F.~Zonzini, M.~M. Malatesta, D.~Bogomolov, N.~Testoni, A.~Marzani, and
  L.~De~Marchi, ``Vibration-based {SHM} with upscalable and low-cost sensor
  networks,'' {\em IEEE Trans. Instrum. Meas.}, vol.~69, no.~10,
  pp.~7990--7998, Mar. 2020.

\bibitem{sarwar2020multimetric}
M.~Z. Sarwar, M.~R. Saleem, J.-W. Park, D.-S. Moon, and D.~J. Kim,
  ``Multimetric event-driven system for long-term wireless sensor operation for
  {SHM} applications,'' {\em IEEE Sens. J.}, vol.~20, no.~10, pp.~5350--5359,
  Feb. 2020.

\bibitem{9547763}
H.~V. Dang, M.~Tatipamula, and H.~X. Nguyen, ``Cloud-based digital twinning for
  structural health monitoring using deep learning,'' {\em IEEE Trans. Ind.
  Inf.}, vol.~18, no.~6, pp.~3820--3830, Sept. 2022.

\bibitem{liu2020internet}
Y.~Liu, C.~Wu, H.~Wu, and Y.~Zhang, ``Internet of things based construction and
  health monitoring of high-pier long-span continuous rigid frame bridge,'' in
  {\em Proc. 2020 Int. Conf. iThings, GreenCom, CPSCom, SmartData,
  Cybermatics}, pp.~276--281, Nov. 2020.

\bibitem{yang2022research}
D.~Yang, L.~Wang, and Y.~Zhang, ``Research on the application of computer big
  data technology in the health monitoring of the bridge body of cross-river
  bridge,'' in {\em Proc. 2022 IEEE Asia-Pac. Conf. IPEC}, pp.~1516--1520, May.
  2022.

\bibitem{shi2024board}
J.~Shi, H.~Shi, Z.~Wu, and J.~Li, ``An on-board damage detection method for
  heavy-haul railway bridge based on sensitivity analysis of bogie responses,''
  {\em IEEE Sens. J.}, vol.~24, no.~4, pp.~4642--4655, Jan. 2024.

\bibitem{zinno2022artificial}
R.~Zinno, S.~S. Haghshenas, G.~Guido, and A.~VItale, ``Artificial intelligence
  and structural health monitoring of bridges: A review of the
  state-of-the-art,'' {\em IEEE Access}, vol.~10, pp.~88058--88078, Aug. 2022.

\bibitem{fawad2024integration}
M.~Fawad, M.~Salamak, M.~U. Hanif, K.~Koris, M.~Ahsan, H.~Rahman, M.~Gerges,
  and M.~M. Salah, ``Integration of bridge health monitoring system with
  augmented reality application developed using 3{D} game engine--case study,''
  {\em IEEE Access}, vol.~12, pp.~16963--16974, Jan. 2024.

\bibitem{zalt2007evaluating}
A.~Zalt, V.~Meganathan, S.~Yehia, O.~Abudayyeh, and I.~Abdel-Qader,
  ``Evaluating sensors for bridge health monitoring,'' in {\em Proc. 2007 IEEE
  Int. Conf. Electro/Inf. Technol.}, pp.~368--372, May. 2007.

\bibitem{bruschetta2013fusion}
F.~Bruschetta, D.~Zonta, C.~Cappello, R.~Zandonini, M.~Pozzi, B.~Glisic,
  D.~Inaudi, D.~Posenato, M.~Wang, and Y.~Zhao, ``Fusion of monitoring data
  from cable-stayed bridge,'' in {\em Proc. 2013 IEEE Workshop Environ. Energy
  Struct. Monit. Syst.}, pp.~1--6, Sept. 2013.

\bibitem{koganezawa2024vibration}
S.~Koganezawa, S.~Terai, H.~Tani, R.~Lu, and S.~Kawada, ``Vibration based
  scour-detection for bridge-piers using a self-powered magnetostrictive
  vibration sensor,'' {\em IEEE Sens. Lett.}, Jul. 2024.

\bibitem{zhou2024novel}
Y.~Zhou, M.~Ding, Z.~Zhu, S.~Su, H.~Chi, and Y.~Cao, ``A novel coplanar
  capacitive micro-displacement sensor combining microfluidic channels for
  bridge health monitoring,'' in {\em Proc. 2024 IEEE Int. Conf. CIS, RAM},
  pp.~75--80, Aug. 2024.

\bibitem{ferguson2024systematic}
A.~J. Ferguson, D.~Hester, and R.~Woods, ``A systematic approach to filter
  specification for measuring quasi-static bridge rotation under moving loads
  using {DC} accelerometers,'' {\em IEEE Access}, vol.~12, pp.~67425--67437,
  May. 2024.

\bibitem{michel2023assessing}
C.~Michel and S.~Keller, ``Assessing important uncertainty influences of
  ground-based radar for bridge monitoring,'' {\em IEEE Geosci. Remote Sens.
  Lett.}, vol.~21, pp.~1--5, Dec. 2023.

\bibitem{pramudita2023fmcw}
A.~A. Pramudita, D.-B. Lin, A.~A. Dhiyani, H.~H. Ryanu, T.~Adiprabowo, and
  E.~A. Yudha, ``{FMCW} radar for noncontact bridge structure displacement
  estimation,'' {\em IEEE Trans. Instrum. Meas.}, vol.~72, pp.~1--14, Jul.
  2023.

\bibitem{sun2024deformation}
C.~Sun, G.~Li, Z.~Hu, Y.~Wang, Z.~Dong, and T.~Zeng, ``Deformation monitoring
  of truss structure bridge with time-series {I}n{SAR} analysis,'' {\em IEEE J.
  Sel. Top. Appl. Earth Obs. Remote Sens.}, Dec. 2024.

\bibitem{wang20242}
B.~Wang, W.~Li, C.~Zhao, Q.~Zhang, G.~Li, X.~Liu, B.~Yan, X.~Cai, J.~Zhang, and
  S.~Zheng, ``{L}2-norm quasi 3-{D} phase unwrapping assisted multitemporal
  {I}n{SAR} deformation dynamic monitoring for the cross-sea bridge,'' {\em
  IEEE J. Sel. Top. Appl. Earth Obs. Remote Sens.}, Oct. 2024.

\bibitem{wang2021review}
X.~Wang, Q.~Zhao, R.~Xi, C.~Li, G.~Li, and L.~Li, ``Review of bridge structural
  health monitoring based on {GNSS}: From displacement monitoring to dynamic
  characteristic identification,'' {\em IEEE Access}, vol.~9, pp.~80043--80065,
  May. 2021.

\bibitem{civera2021computer}
M.~Civera, L.~Z. Fragonara, and C.~Surace, ``A computer vision-based approach
  for non-contact modal analysis and finite element model updating,'' in {\em
  Eur. Workshop Struct. Health Monit., Spec. Collect.}, vol.~1, pp.~481--493,
  Springer, Jan. 2021.

\bibitem{alonso2024contribution}
A.~Alonso-D{\'\i}az, M.~Solla, A.~Elseicy, and J.~L. Rodr{\'\i}guez,
  ``Contribution of the {MT}-{I}nsar technique for the monitoring of transport
  infrastructures,'' in {\em Proc. IGARSS 2024}, pp.~11141--11145, Jul. 2024.

\bibitem{hu2024vehicle}
M.~Hu, J.~Li, L.~Luo, and J.~Yue, ``Vehicle overload warning method based on
  {GNSS} displacement signals of long-span bridges,'' in {\em Proc. 2024 3rd
  Int. Symp. ISSET}, pp.~227--231, Aug. 2024.

\bibitem{topal2024filtering}
G.~Oku~Topal and B.~Akpinar, ``Filtering low-cost {GNSS} measurements to
  determine structural behaviors: The case of davutpaşa pedestrian bridge,''
  {\em IEEE Access}, vol.~12, pp.~101184--101196, Jul. 2024.

\bibitem{liu2022integrated}
F.~Liu, Y.~Cui, C.~Masouros, J.~Xu, T.~X. Han, Y.~C. Eldar, and S.~Buzzi,
  ``Integrated sensing and communications: Toward dual-functional wireless
  networks for 6{G} and beyond,'' {\em IEEE J. Sel. Areas Commun.}, vol.~40,
  no.~6, pp.~1728--1767, Mar. 2022.

\bibitem{giordani2020toward}
M.~Giordani, M.~Polese, M.~Mezzavilla, S.~Rangan, and M.~Zorzi, ``Toward 6{G}
  networks: Use cases and technologies,'' {\em IEEE Commun. Mag.}, vol.~58,
  no.~3, pp.~55--61, Mar. 2020.

\bibitem{de2021convergent}
C.~De~Lima, D.~Belot, R.~Berkvens, A.~Bourdoux, D.~Dardari, M.~Guillaud,
  M.~Isomursu, E.-S. Lohan, Y.~Miao, A.~N. Barreto, {\em et~al.}, ``Convergent
  communication, sensing and localization in 6{G} systems: An overview of
  technologies, opportunities and challenges,'' {\em IEEE Access}, vol.~9,
  pp.~26902--26925, Jan. 2021.

\bibitem{zhong2025resource}
C.~Zhong, H.~Ding, D.~Li, L.~Tang, and Y.-c. Liang, ``Resource allocation for
  dynamic {TDD}-enabled integrated sensing and communication systems,'' {\em
  IEEE Trans. Veh. Technol.}, pp.~1--13, Feb. 2025.

\bibitem{2024arXiv240519925L}
H.~{Luo}, T.~{Zhang}, C.~{Zhao}, Y.~{Wang}, B.~{Lin}, Y.~{Jiang}, D.~{Luo}, and
  F.~{Gao}, ``{Integrated sensing and communications framework for 6{G}
  networks},'' {\em arXiv e-prints}, p.~arXiv:2405.19925, May 2024.

\bibitem{liao2023optimized}
C.~Liao, F.~Wang, and V.~K. Lau, ``Optimized design for {IRS}-assisted
  integrated sensing and communication systems in clutter environments,'' {\em
  IEEE Trans. Commun.}, vol.~71, no.~8, pp.~4721--4734, Jun. 2023.

\bibitem{yang2024application}
J.~Yang, C.-K. Wen, and S.~Jin, ``Application of integrated sensing and
  communication in structural health monitoring,'' in {\em Proc. 2024 33rd
  WOCC}, pp.~128--133, Oct. 2024.

\bibitem{10868647}
D.~Chen, S.~Xia, S.~Chen, Y.~Ma, and Z.~Wang, ``Integrated sensing and
  communication based deformation monitoring in practical 5{G} network,'' in
  {\em Proc. 2024 IEEE Int. Conf. ICSIDP}, pp.~1--6, Nov. 2024.

\bibitem{8288677}
F.~Liu, C.~Masouros, A.~Li, H.~Sun, and L.~Hanzo, ``{MU}-{MIMO} communications
  with {MIMO} radar: From co-existence to joint transmission,'' {\em IEEE
  Trans. Wireless Commun.}, vol.~17, no.~4, pp.~2755--2770, Feb. 2018.

\bibitem{9947033}
Z.~Du, F.~Liu, W.~Yuan, C.~Masouros, Z.~Zhang, S.~Xia, and G.~Caire,
  ``Integrated sensing and communications for {V}2{I} networks: Dynamic
  predictive beamforming for extended vehicle targets,'' {\em IEEE Trans.
  Wireless Commun.}, vol.~22, no.~6, pp.~3612--3627, Jun. 2023.

\bibitem{10147248}
Y.~Xiong, F.~Liu, Y.~Cui, W.~Yuan, T.~X. Han, and G.~Caire, ``On the
  fundamental tradeoff of integrated sensing and communications under gaussian
  channels,'' {\em IEEE Trans. Inf. Theory}, vol.~69, no.~9, pp.~5723--5751,
  Jun. 2023.

\bibitem{8386661}
F.~Liu, L.~Zhou, C.~Masouros, A.~Li, W.~Luo, and A.~Petropulu, ``Toward
  dual-functional radar-communication systems: Optimal waveform design,'' {\em
  IEEE Trans. Signal Process.}, vol.~66, no.~16, pp.~4264--4279, Aug. 2018.

\bibitem{9755276}
Z.~Wei, F.~Liu, C.~Masouros, N.~Su, and A.~P. Petropulu, ``Toward
  multi-functional 6{G} wireless networks: Integrating sensing, communication,
  and security,'' {\em IEEE Commun. Mag.}, vol.~60, no.~4, pp.~65--71, Apr.
  2022.

\bibitem{10625724}
D.~Ma, J.~Wei, Q.~Zhang, Z.~Wei, W.~Jiang, N.~Guo, and S.~Huang, ``Performance
  evaluation of micro-doppler based {UAV} identification using different 5{G}
  frame structures,'' in {\em Proc. 2024 2nd Int. Conf. MICCIS}, pp.~173--179,
  Apr. 2024.

\bibitem{7729680}
H.~Dong, J.~Wang, and Q.~Song, ``The analysis and verification about the update
  rate constraint for the interferometric radar of displacement measurement,''
  in {\em 2016 IEEE Int. Geoscience and Remote Sensing Symposium (IGARSS)},
  pp.~2634--2637, 2016.

\bibitem{ma2023structural}
Z.~Ma, J.~Choi, L.~Yang, and H.~Sohn, ``Structural displacement estimation
  using accelerometer and {FMCW} millimeter wave radar,'' {\em Mechanical
  Systems and Signal Processing}, vol.~182, p.~109582, Jan. 2023.

\bibitem{10854508}
H.~Luo, F.~Gao, F.~Liu, and S.~Jin, ``6{D} motion parameters estimation in
  monostatic integrated sensing and communications system,'' {\em IEEE Trans.
  Commun.}, pp.~1--1, 2025.

\bibitem{10477890}
H.~Luo, Y.~Wang, D.~Luo, J.~Zhao, H.~Wu, S.~Ma, and F.~Gao, ``Integrated
  sensing and communications in clutter environment,'' {\em IEEE Trans.
  Wireless Commun.}, vol.~23, no.~9, pp.~10941--10956, Mar. 2024.

\bibitem{9945983}
F.~Dong, F.~Liu, Y.~Cui, W.~Wang, K.~Han, and Z.~Wang, ``Sensing as a service
  in 6{G} perceptive networks: A unified framework for {ISAC} resource
  allocation,'' {\em IEEE Trans. Wireless Commun.}, vol.~22, no.~5,
  pp.~3522--3536, Nov. 2023.

\bibitem{5776640}
C.~Sturm and W.~Wiesbeck, ``Waveform design and signal processing aspects for
  fusion of wireless communications and radar sensing,'' {\em Proc. IEEE},
  vol.~99, no.~7, pp.~1236--1259, Jul. 2011.

\end{thebibliography}

	\vfill
	
\end{document}